\documentclass[12pt, draftclsnofoot, onecolumn]{IEEEtran}
\makeatletter
\def\ps@headings{%
\def\@oddhead{\mbox{}\scriptsize\rightmark \hfil \thepage}%
\def\@evenhead{\scriptsize\thepage \hfil \leftmark\mbox{}}%
\def\@oddfoot{}%
\def\@evenfoot{}}
\makeatother 
\IEEEoverridecommandlockouts

\usepackage{graphicx}
\usepackage{amsmath,amssymb}

\usepackage{algorithm}
\usepackage{algpseudocode}
\usepackage{amsmath}
\usepackage{graphics}
\usepackage{epsfig}
\usepackage{color}
\usepackage{subfigure}

\newtheorem{remark}{\underline{Remark}}[section]

\newtheorem{lemma}{Lemma}
\usepackage[numbers,sort&compress]{natbib}
\usepackage{flafter}

\newcommand{\mv}[1]{\mbox{\boldmath{$ #1 $}}}
\begin{document}
%
\title{Joint 3D Maneuver and Power Adaptation for Secure UAV Communication with CoMP Reception}
\author{\IEEEauthorblockN{Jianping~Yao and Jie~Xu}
\vspace{-3em}
\thanks
{
Part of this paper has been presented at the IEEE Global Communications Conference (GLOBECOM), Waikoloa, Hawaii, USA, December 9-13, 2019 \cite{yao3D2019}.
}
\thanks
{
J. Yao is with the School of Information Engineering, Guangdong University of Technology, Guangzhou 510006, China
(e-mail: yaojp@gdut.edu.cn).
}
\thanks{J. Xu is with the Future Network of Intelligence Institute (FNii) and the School of Science and Engineering, the Chinese University of Hong Kong, Shenzhen, Shenzhen 518172, China. He was with the School of Information Engineering, Guangdong University of Technology, Guangzhou 510006, China (e-mail: jiexu.ustc@gmail.com). J. Xu is the corresponding author.}
}

\maketitle

\begin{abstract}
This paper studies a secrecy unmanned aerial vehicle (UAV) communication system with coordinated multi-point (CoMP) reception, in which one UAV sends confidential messages to a set of cooperative ground receivers (GRs), in the presence of several suspicious eavesdroppers.
In particular, we consider two types of eavesdroppers that are non-colluding and colluding, respectively. Under this setup, we exploit the UAV's maneuver in three dimensional (3D) space together with transmit power adaptation for optimizing the secrecy communication performance.
First, we consider the quasi-stationary UAV scenario, in which the UAV is placed at a fixed but optimizable location during the communication period.
In this scenario, we jointly optimize the UAV's 3D placement and transmit power control to maximize the secrecy rate.
Under both non-colluding and colluding eavesdroppers, we obtain the optimal solutions to the joint 3D placement and transmit power control problems in well structures.
Next, we consider the mobile UAV scenario, in which the UAV has a mission to fly from an initial location to a final location during the communication period.
In this scenario, we jointly optimize the UAV's 3D trajectory and transmit power allocation to maximize the average secrecy rate during the whole communication period.
To deal with the difficult joint 3D trajectory and transmit power allocation problems, we present alternating-optimization-based approaches to obtain high-quality solutions. Finally, we provide numerical results to validate the performance of our proposed designs. It is shown that due to the consideration of CoMP reception, our proposed design with 3D maneuver significantly outperforms the conventional design with two dimensional (2D) (horizontal) maneuver only, by exploiting the additional degrees of freedom in altitudes.
It is also shown that the non-colluding and colluding eavesdroppers lead to distinct 3D UAV maneuver behaviors, e.g., under colluding eavesdroppers, the UAV should fly farther apart from them (than that under the non-colluding ones) for avoiding their collaborative interception.
\end{abstract}


\section{Introduction}
Unmanned aerial vehicles (UAVs) are envisioned to play an important role in beyond fifth-generation (B5G) and sixth-generation (6G) cellular networks \cite{ZengAccessing2019}.
On one hand, UAVs with certain missions such as cargo delivery and aerial inspection can be connected seamlessly with on-ground cellular base stations (BSs), in order to increase the communication and operation range, and enhance the quality of service \cite{ZengCellular2019,ZhangCellular2019,MozaffariBeyond2019,MenouarUAV2017,XiaoEnabling2016}.
On the other hand, UAVs can also be employed as aerial wireless platforms (e.g., BSs, relays, and wireless chargers) in the sky to provide flexible and on-demand wireless services to ground users, with both improved transmission efficiency and enhanced coverage (see, e.g., \cite{ZengWireless2016,XuUAV2018,LiPlacement2018,XieThroughput2019,EsrafilianLearning2019,ChenEfficient2019,BerghLTE2016,YalinizThe2016} and the references therein). With the emergence of cellular-connected UAVs and UAV-assisted wireless transmissions, various technical opportunities and challenges have been imposed. First, UAVs in the sky normally have strong line-of-sight (LoS) air-to-ground (A2G) links with on-ground nodes. This leads to better A2G communication quality but also stronger A2G interference. Next, UAVs have high mobility in the three-dimensional (3D) space. This makes the mobility management a difficult task, but also provides a new design degree of freedom for improving the communication performance via UAV placement or trajectory control.

Among others, the security issue is another critical challenge faced in UAV communications. Due to the broadcast nature of wireless channels and the existence of strong LoS components over A2G links, the transmitted signals from UAVs in the sky are more vulnerable to be eavesdropped by suspicious nodes on the ground than conventional terrestrial communications.
Physical layer security has been recognized as a viable solution to protect wireless communications against eavesdropping attacks (see, e.g., \cite{YaoSecure2016,YanSecret2018} and the references therein).
Different from conventional cryptology-based security technology, physical layer security is able to provide perfect security for wireless communication systems from an information theoretical perspective. Therefore, to provide secure UAV communications, it is emerging and of great importance to conduct research on using physical layer security for securing UAV communications.

In the literature, there have been various prior works investigating the integration of physical layer security in UAV communications and networks (see, e.g., \cite{Yao2019secrecy,ZhuSecrecy2018,TangSecrecy2019,CuiRobust2018,ZhangSecuring2018,WangImproving2017,ZhouSecrecy2017,LiRobust2018,BaiEnergy2019,LeeUAV2018,ZhongSecure2018,CaiDual2018,ZhaoCaching2018,ZhouUAV2019}).
In general, these works can be roughly classified into two categories that considered the network-level performance analysis for large-scale random UAV networks via stochastic geometry \cite{Yao2019secrecy,ZhuSecrecy2018,TangSecrecy2019}, and the link-level performance optimization via UAV maneuver design and wireless resource allocation \cite{CuiRobust2018,ZhangSecuring2018,WangImproving2017,ZhouSecrecy2017,LiRobust2018,BaiEnergy2019,LeeUAV2018,ZhongSecure2018,CaiDual2018,ZhaoCaching2018,ZhouUAV2019}, respectively.
In particular, we focus on the UAV maneuver design for optimizing the link-level performance. Intuitively speaking, by exploiting the controllable mobility in 3D space, legitimate UAV transmitters can fly close to intended ground receivers (GRs) to improve the legitimate channel quality and move far away from suspicious eavesdroppers to prevent the information leakage, thus improving the security of legitimate transmission.
For instance, \cite{CuiRobust2018,ZhangSecuring2018,WangImproving2017,ZhouSecrecy2017,LiRobust2018,BaiEnergy2019} considered the scenario with one UAV communicating with one GR or acting as a mobile relay in the presence of one eavesdropper, in which the UAV's trajectory and power allocation are jointly optimized for maximizing the secrecy rate.
\cite{LeeUAV2018,ZhongSecure2018,CaiDual2018,ZhaoCaching2018,ZhouUAV2019} then considered the scenario when one UAV's legitimate communication is assisted by the other UAV's cooperative jamming, in which the communicating and jamming UAVs jointly design their trajectories and power allocations for further improving the secrecy performance.
Nevertheless, these existing works on secrecy UAV communications mainly focused on the scenario with each GR independently decoding the respective messages without cooperation, under which the UAV is considered to fly at a fixed altitude with only two-dimensional (2D) horizontal trajectory optimized.

\begin{figure}[!t]
\centering
\includegraphics[width=6.8cm]{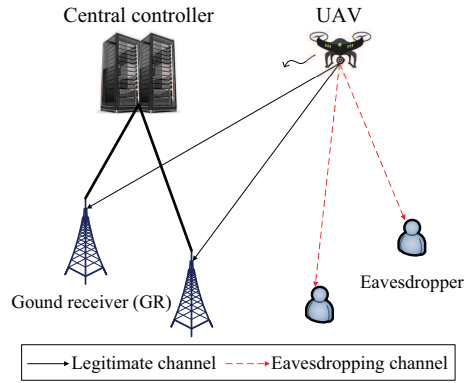}
\caption{Illustration of the secrecy UAV communication system with CoMP reception, in the presence of several suspicious eavesdroppers.}
\label{Fig:system model}
\end{figure}

Recently, the coordinated multi-point (CoMP) transmission/reception has been recognized as a promising technique in wireless networks \cite{SawahashiCoordinated2010,CheckoCloud2015,SimeoneCloud2016,LiuCoMP2019}, which enables symbol-level cooperation among geographically distributed nodes (such as BSs) to increase the communication performance via utilizing the inter-cell interference. It is well established that CoMP is able to significantly enhance the communication reliability and increase the data-rate throughput for both cell-center and cell-edge users \cite{SawahashiCoordinated2010,CheckoCloud2015,SimeoneCloud2016,LiuCoMP2019}, and also increase the secrecy communication performance \cite{NgSecure2015}.
Therefore, it is expected that the exploitation of CoMP can also enhance the performance of secrecy UAV communications. Nevertheless, how to optimally design the UAV maneuver for secrecy UAV communication under CoMP has not been studied in the literature yet. This thus motivates our investigation in this work.

In this paper, we consider the secrecy UAV communication with CoMP reception over a finite communication period, in which one UAV communicates with multiple cooperative legitimate GRs, in the presence of multiple suspicious eavesdroppers.
With CoMP, these GRs are enabled to cooperatively decode the legitimate messages sent from the UAV to defend against the eavesdropping attack.
In particular, we consider two types of eavesdroppers that are non-colluding and colluding, respectively.
Different from previous works that focused on 2D (horizontal) maneuver, we further exploit the vertical maneuver (or equivalently, altitude) via 3D placement/trajectory design, together with the transmit power adaptation, to facilitate the secure UAV communication.
Under this setup, the main results of this paper are summarized as follows.
\begin{itemize}
\item First, we consider the quasi-stationary UAV scenario, in which the UAV is placed at a fixed but optimizable location during the communication period.
In this scenario, we jointly optimize the UAV's 3D placement and transmit power control to maximize the secrecy rate, subject to the UAV's minimum/maximum altitude and maximum transmit power constraints.
Although the formulated secrecy rate maximization problems under non-colluding and colluding eavesdroppers are both non-convex, we propose efficient methods to obtain the optimal solutions in well structures.
\item Next, we consider the mobile UAV scenario, in which the UAV has a mission to fly from an initial location to a final location during the communication period. In this scenario, we jointly optimize the UAV's 3D trajectory and transmit power allocation to maximize the average secrecy rate during the whole communication period, subject to its maximum flight speed and maximum transmit power constraints.
Due to the coupling between transmit power and trajectory variables, the formulated secrecy rate maximization problems are non-convex, which are very difficult to be solved optimally.
To tackle this difficulty, we propose alternating-optimization-based approaches to optimize the transmit power allocation and trajectory design alternately, by using the convex optimization and successive convex approximation (SCA) techniques, respectively.
\item Finally, we provide numerical results to validate the performance of our proposed designs. It is shown that the joint 3D maneuver and transmit power optimization greatly enhances the secrecy communication performance under our setup, as compared to other benchmark schemes with e.g. 2D maneuver optimization only.
It is also shown that the 3D maneuver behaviors under non-colluding and colluding eavesdroppers are distinct, where the UAV generally should fly farther apart from the eavesdroppers (for avoiding the collaborative interception) if they are colluding, but the UAV may fly relatively closer to them (to combat against their individual eavesdropping) if they are non-colluding.
\end{itemize}

It is worth pointing out that in the literature, the 3D placement and/or trajectory design has been investigated in UAV communications under other setups \cite{KalantariOn2016,Alzenad3D2017,SunOptimal2019,HuangCognitive2019} (instead of secrecy UAV communications with CoMP reception in this paper).
For example, \cite{KalantariOn2016,Alzenad3D2017} optimized the UAVs' 3D placement design for serving ground users more cost-effectively.
\cite{SunOptimal2019} investigated the joint design of 3D trajectory and resource allocation for maximizing the system sum throughput in solar-powered UAV communication systems, in which the UAV can adjust its vertical locations for balancing between the harvested solar power level versus the communication channel quality.
\cite{HuangCognitive2019} considered the spectrum sharing between a cognitive UAV communication system and a coexisting primary terrestrial communications, in which the 3D UAV trajectory is controlled to balance between the cognitive communication performance versus the A2G interference towards the primary terrestrial system.
To our best knowledge, our proposed 3D maneuver design for secure UAV communications with CoMP reception is new and has not been studied yet.

The remainder of this paper is organized as follows. Section \uppercase\expandafter{\romannumeral2} presents the system model of our considered CoMP reception-enabled secrecy UAV communication system.
Section \uppercase\expandafter{\romannumeral3} optimizes the 3D placement and transmit power control to maximize the secrecy rate in the quasi-stationary UAV scenario.
Section \uppercase\expandafter{\romannumeral4} optimizes the 3D trajectory and transmit power allocation to maximize the average secrecy rate in the mobile UAV scenario.
Section \uppercase\expandafter{\romannumeral5} presents numerical results to validate the secrecy communication performance of our proposed designs.
Finally, Section \uppercase\expandafter{\romannumeral6} concludes this paper.
%
%

\begin{table}[!htbp]
\vspace{-3.0em}
\centering
\caption{List of Notations}\vspace{-2.0em}
\label{tab:List}
\begin{tabular}{ll}
$K$                              &Number of GRs\\
$\mathcal{K}$                    &Set of GRs\\
$J$                              &Number of eavesdroppers\\
$\mathcal{J}$                    &Set of eavesdroppers\\
$T$                              &Duration of the communication period of interest\\
$N$                              &Number of time slots\\
$\mathcal{N}$                    &Set of time slots\\
$t_s$                            &Duration of each time slot\\
$\mv w_{bk}$                     &Horizontal location of GR $k$\\
$\mv w_{ej}$                     &Horizontal location of eavesdropper $j$\\
$\beta_0$                        &Channel power gain at a reference distance of 1 m\\
$d_{bk}$                         &Distance from UAV to GR $k$\\
$d_{ej}$                         &Distance from UAV to eavesdropper $j$\\
$(x,y,z)$                        &3D location of UAV\\
$\mv{q}$                         &Horizontal location of UAV\\
$z_{\text{min}}$                 &Minimum flight altitude of UAV\\
$z_{\text{max}}$                 &Maximum flight altitude of UAV\\
$\tilde{g}_{bk}$                 &Channel power gain from UAV to GR $k$\\
$\tilde{g}_{ej}$                 &Channel power gain from UAV to eavesdropper $j$\\
$p$                              &Transmit power of UAV\\
$P$                              &Maximum power level of UAV in quasi-stationary UAV scenario\\
$P_\text{peak}$                  &Maximum peak transmit power of UAV in mobile UAV scenario\\
$P_\text{ave}$                   &Maximum average transmit power of UAV in mobile UAV scenario\\
$s$                              &Transmitted signal by UAV\\
$y_{bk}$                         &Received signal at GR $k$\\
$y_{ej}$                         &Received signal at eavesdropper $j$\\
$n_{bk}$                         &AWGN at receiver of each individual GR $k$\\
$n_{ej}$                         &AWGN at receiver of each individual eavesdropper $j$\\
$\gamma_{bk}$                    &Received SNR at GR $k$\\
$\gamma_{ej}$                    &Received SNR at eavesdropper $j$\\
$g_{bk}$                         &Channel-power-to-noise ratio from UAV to GR $k$\\
$g_{ej}$                         &Channel-power-to-noise ratio from UAV to eavesdropper $j$\\
$\gamma_b$                       &Received SNR at GRs with CoMP reception\\
$\gamma_e^{(\text{\uppercase\expandafter{\romannumeral1}})}$ &Received SNR at non-colluding eavesdroppers\\
$\gamma_e^{(\text{\uppercase\expandafter{\romannumeral2}})}$ &Received SNR at colluding eavesdroppers\\
$R^{(\text{\uppercase\expandafter{\romannumeral1}})}$ &Secrecy rate under non-colluding eavesdroppers\\
$R^{(\text{\uppercase\expandafter{\romannumeral2}})}$ &Secrecy rate under colluding eavesdroppers\\
$\tilde{V}$                      &Maximum horizontal flight speed of UAV\\
$V$                              &Maximum horizontal displacement of UAV\\
$\tilde{V}_{\text{up}}$          &Maximum vertical ascending speed of UAV\\
$\tilde{V}_{\text{down}}$        &Maximum vertical descending speed of UAV\\
${V}_{\text{up}}$                &Maximum vertical ascending displacement of UAV\\
${V}_{\text{down}}$              &Maximum vertical descending displacement of UAV
\end{tabular}
\end{table}

\section{System Model}
In this work, we consider the secrecy UAV communication system as shown in Fig. \ref{Fig:system model}, in which one UAV communicates with $K$ GRs in the presence of $J$ suspicious eavesdroppers on the ground.
This may practically correspond to the cellular-connected UAV scenario with the GRs being ground BSs that can cooperate in CoMP reception to jointly decode the legitimate messages sent from the UAV.
We focus on a particular UAV communication period with duration $T$ in second (s).
Without loss of generality, we consider a 3D Cartesian coordinate system with the fixed horizontal location of GR $k$ being $\mv w_{bk}\in\mathbb{R}^2$, $k\in\mathcal{K}\triangleq\{1, \ldots,K\}$, and that of eavesdropper $j$ being $\mv w_{ej}\in\mathbb{R}^2$, $j\in\mathcal{J}\triangleq\{1, \ldots,J\}$.
It is assumed that the UAV perfectly knows the locations of both GRs and eavesdroppers {\it a-priori}.\footnote{Notice that our results are extendible to the case when the UAV only has partial knowledge of the GRs' and/or eavesdroppers' locations, by e.g. using the robust optimization techniques based on bounded location error models (see, e.g., \cite{ZhongSecure2018}).}
This assumption is made for the purpose of facilitating the joint maneuver and power control design, and gaining essential design insights. In practice, the GRs' locations can be acquired by the UAV via the GRs reporting such information; while the eavesdroppers' locations may be obtained by the UAV via monitoring the eavesdroppers' emitted signals (if the eavesdroppers belong to a different network entity as the UAV) \cite{KapetanovicPhysical2015}\footnote{Even if the eavesdroppers are passive in reception only, it is possible to detect the passive eavesdropping from the local oscillator power leaked from the eavesdroppers' RF front end \cite{Mukherjee2012}.}, or via a centralized network controller reporting such information (if the eavesdroppers belong to the same network entity as the UAV) \cite{LiangSecure2008}.
For convenience, we summarize the main notations used in this paper in Table \ref{tab:List}.
In the following, we consider two scenarios with quasi-stationary and mobile UAVs, respectively.

\subsection{Quasi-Stationary UAV Scenario}
First, we consider the quasi-stationary UAV scenario, in which the UAV is placed at a fixed but optimizable location $(x,y,z)$ over the whole duration-$T$ communication period.
For notational convenience, let $\mv{q}=(x,y)$ denote the horizontal location of the UAV, and $z$ denote its vertical location or altitude, respectively.
Let $z_{\text{min}}$ and $z_{\text{max}}$ denote the minimally and maximally allowed UAV flight altitudes, respectively, with $z_{\text{min}}\leq z \leq z_{\text{max}}$, which are set for safety reasons based on certain regulations.
As A2G channels from the UAV to ground nodes (both GRs and eavesdroppers) normally have strong LoS links, we consider a LoS channel model with a generic path loss exponent $\alpha$ with $\alpha \in [2,4]$ in general.
Accordingly, the channel power gain from the UAV to each GR $k\in\mathcal{K}$ is given by
\begin{align}
\tilde{g}_{bk}\left(\mv{q},z\right)=\frac{\beta_0}{d_{bk}^{\alpha}\left(\mv{q},z\right)}=\frac{\beta_0}{\left(||\mv{q}-\mv{w}_{bk}||^2 + z^2\right)^{\frac{\alpha}{2}}},
\label{Channel_Power_GR_static}
\end{align}
where $\beta_0$ denotes the channel power gain at the reference distance of $1$ meter (m) and $d_{bk}\left(\mv{q},z\right)$ denotes the distance from the UAV to GR $k$.
Similarly, the channel power gain from the UAV to eavesdropper $j\in\mathcal{J}$ is
\begin{align}
\tilde{g}_{ej}\left(\mv{q},z\right)=\frac{\beta_0}{d_{ej}^{\alpha}\left(\mv{q},z\right)}=\frac{\beta_0}{\left(||\mv{q}-\mv{w}_{ej}||^2 + z^2\right)^{\frac{\alpha}{2}}},
\label{Channel_Power_EVE_static}
\end{align}
where $d_{ej}\left(\mv{q},z\right)$ denotes the distance from the UAV to eavesdropper $j$.
Let $p \ge 0$ denote the transmit power by the UAV, which is subject to a maximum power level $P$.
Accordingly, we have
\begin{align}
0\le p \le P. \label{Power_peak_static}
\end{align}

Next, we consider the secure communication from the UAV to the $K$ legitimate GRs.
Let $s$ denote the UAV's transmitted signal that is a circularly symmetric complex Gaussian (CSCG) random variable with zero mean and unit variance, i.e., $s \sim \mathcal{CN}(0,1)$.
In this case, the received signals at each GR $k\in\mathcal{K}$ and eavesdropper $j\in\mathcal{J}$ are respectively given as
\begin{align}
y_{bk} =\sqrt{\tilde{g}_{bk}\left(\mv{q},z\right)p}s+n_{bk},\\
y_{ej} =\sqrt{\tilde{g}_{ej}\left(\mv{q},z\right)p}s+n_{ej},
\end{align}
where $n_{bk}$ and $n_{ej}$ denote the additive white Gaussian noise (AWGN) at the receivers of GR $k$ and eavesdropper $j$, respectively, each with zero mean and variance $\sigma^2$, i.e., ${n_{bk}}\sim\mathcal{CN}(0,\sigma^2)$, ${n_{ej}}\sim\mathcal{CN}(0,\sigma^2)$.
Then, the received signal-to-noise ratios (SNRs) at each individual GR $k\in\mathcal{K}$ and eavesdropper $j\in\mathcal{J}$ are respectively given as
\begin{align}
\gamma_{bk}\left(\mv{q},z,p\right)
=\tilde{g}_{bk}\left(\mv{q},z\right)p/\sigma^2 = g_{bk}\left(\mv{q},z\right)p,\label{SNR_GR_static}\\
\gamma_{ej}\left(\mv{q},z,p\right)={\tilde{g}_{ej}\left(\mv{q},z\right)p/\sigma^2}={g_{ej}\left(\mv{q},z\right)p},\label{SNR_EVE_static}
\end{align}
where for notational convenience, $g_{bk}\left(\mv{q},z\right) = \tilde{g}_{bk}\left(\mv{q},z\right)/\sigma^2$ and $g_{ej}\left(\mv{q},z\right) = \tilde{g}_{ej}\left(\mv{q},z\right)/\sigma^2$ are defined as the channel-power-to-noise ratios from the UAV to GR $k$ and eavesdropper $j$, respectively.

With CoMP reception, the $K$ GRs jointly decode the received legitimate message $s$ via maximal ratio combining (MRC). Accordingly, the received SNR at the GRs is given as
\begin{align}
\gamma_b\left(\mv{q},z,p\right) &=\sum\limits_{k\in\mathcal{K}}\gamma_{bk}\left(\mv{q},z,p\right) = \sum\limits_{k\in\mathcal{K}}g_{bk}\left(\mv{q},z\right)p.
\label{SNRs_GRs_static}
\end{align}

In particular, we consider two types of eavesdroppers that are non-colluding and colluding, namely Type-I and Type-II eavesdroppers, respectively.
First, consider the non-colluding eavesdroppers, which can only intercept/decode the confidential message from the UAV individually. In this case, the received SNR at the non-colluding eavesdroppers is limited by the one with the strongest signal, which is given by
\begin{align}
\gamma_e^{(\text{\uppercase\expandafter{\romannumeral1}})}\left(\mv{q},z,p\right)&=\max\limits_{j\in\mathcal{J}}\gamma_{ej}\left(\mv{q},z,p\right)=\max\limits_{j\in\mathcal{J}}{g_{ej}\left(\mv{q},z\right)p}.
\label{SNRs_EVE_Colluding_static}
\end{align}
As a result, the secrecy rate from the UAV to the $K$ GRs (in bits-per-second-per-Hertz, bps/Hz) for the case of Type-{\uppercase\expandafter{\romannumeral1}} eavesdroppers is given by \cite{ZhongSecure2018}
\begin{align}\label{Rate_static_noncolluding}
\nonumber R^{(\text{\uppercase\expandafter{\romannumeral1}})}\left(\mv{q},z,p\right)
&=\left[\log_2\left(1+\gamma_b\left(\mv{q},z,p\right)\right)-\log_2\left(1+\gamma_e^{(\text{\uppercase\expandafter{\romannumeral1}})}\left(\mv{q},z,p\right)\right)\right]^+,\\
&= \left[\log_2\left(1+\sum\limits_{k\in\mathcal{K}}g_{bk}\left(\mv{q},z\right)p\right)-\log_2\left(1+\max\limits_{j\in\mathcal{J}}{g_{ej}\left(\mv{q},z\right)p}\right)\right]^+,
\end{align}
where $[u]^+\triangleq \max(u,0)$.
Next, consider the colluding eavesdroppers, which can cooperatively intercept/decode the confidential message $s$ from the UAV by combining their intercepted signals. Hence, by using the MRC, the received SNR at the colluding eavesdroppers is equivalent to
\begin{align}
\gamma_e^{(\text{\uppercase\expandafter{\romannumeral2}})}\left(\mv{q},z,p\right)=\sum\limits_{j\in\mathcal{J}}\gamma_{ej}\left(\mv{q},z,p\right)=\sum\limits_{j\in\mathcal{J}}{g_{ej}\left(\mv{q},z\right)p}.
\label{SNRs_EVE_Noncolluding_static}
\end{align}
As a result, the secrecy rate from the UAV to the $K$ GRs (in bps/Hz) for the case of Type-{\uppercase\expandafter{\romannumeral2}} eavesdroppers is given by \cite{ZhongSecure2018}
\begin{align}\label{Rate_static_colluding}
\nonumber R^{(\text{\uppercase\expandafter{\romannumeral2}})}\left(\mv{q},z,p\right)
&=\left[\log_2\left(1+\gamma_b\left(\mv{q},z,p\right)\right)-\log_2\left(1+\gamma_e^{(\text{\uppercase\expandafter{\romannumeral2}})}\left(\mv{q},z,p\right)\right)\right]^+,\\
&= \left[\log_2\left(1+\sum\limits_{k\in\mathcal{K}}g_{bk}\left(\mv{q},z\right)p\right)-\log_2\left(1+\sum\limits_{j\in\mathcal{J}}{g_{ej}\left(\mv{q},z\right)p}\right)\right]^+.
\end{align}

In the quasi-stationary UAV scenario, our objective is to jointly optimize the 3D UAV placement $\mv{q}$ and $z$ and the transmit power control $p$, to maximize the secrecy rate from the UAV to GRs (i.e., $R^{(i)}\left(\mv{q},z,p\right), i\in\{\text{\uppercase\expandafter{\romannumeral1}},\text{\uppercase\expandafter{\romannumeral2}}\}$), subject to the UAV's minimum/maximum altitude constraints, as well as the maximum transmit power constraint. For the case with Type-$i$ eavesdroppers, $i\in\{\text{\uppercase\expandafter{\romannumeral1}},\text{\uppercase\expandafter{\romannumeral2}}\}$, the secrecy rate maximization problem is formulated as
\begin{align}
\nonumber(\text{P1-}i):&\max \limits_{\mv{q},z,p}R^{(i)}\left(\mv{q},z,p\right)\\
\nonumber ~\text{s.t.}~ &z_{\text{min}}\leq z \leq z_{\text{max}}\\
\nonumber&0\le p \le P,
\end{align}
where each problem (P1-$i$) is for Type-$i$ eavesdroppers, $i\in\{\text{\uppercase\expandafter{\romannumeral1}},\text{\uppercase\expandafter{\romannumeral2}}\}$.
Note that the objective functions in both problems $(\text{P1-{\uppercase\expandafter{\romannumeral1}}})$ and $(\text{P1-{\uppercase\expandafter{\romannumeral2}}})$ are non-smooth (due to the operator $[\cdot]^+$) and non-concave, with variables $\mv{q}$, $z$, and $p$ coupled.
Therefore, problems $(\text{P1-{\uppercase\expandafter{\romannumeral1}}})$ and $(\text{P1-{\uppercase\expandafter{\romannumeral2}}})$ are both non-convex and generally difficult to be optimally solved.

\subsection{Mobile UAV Scenario}
Next, we consider the mobile UAV scenario, in which the UAV has a mission to fly from an initial location to a final location during the communication period with duration $T$.
In this case, we discretize the communication period into $N$ time slots each with equal duration $t_s = T/N$. Let $\mathcal{N}\triangleq\{1, \ldots,N\}$ denote the set of time slots.
Let $(x[n],y[n],z[n])$ denote the UAV's time-varying 3D location at time slot $n\in\mathcal{N}$, where $\mv{q}[n]=(x[n],y[n])$ denotes its horizontal location, and $z[n]$ denotes its vertical location or altitude.
Also, suppose that $\mv{q}[0]$ and $\mv{q}[N+1]$ denote the UAV's pre-determined initial and final horizontal locations, and $z[0]$ and $z[N+1]$ denote the corresponding altitudes, respectively.
Let $\tilde{V}$ denote the UAV's maximum horizontal speed in meters/second (m/s), and $V=\tilde{V} t_s$ denote the maximum horizontal displacement of the UAV between two consecutive slots.
It then follows that
\begin{align}\label{Traj_constraint}
\|\mv{q}[n+1]-\mv{q}[n]\|\leq V, \forall n\in \{0\}\cup\mathcal{N},
\end{align}
where $\|\cdot\|$ denotes the Euclidean norm.
Let $\tilde{V}_{\text{up}}$ and $\tilde{V}_{\text{down}}$ denote the maximum vertical ascending and descending speeds, respectively.
Then, we have $V_{\text{up}}=\tilde{V}_{\text{up}} t_s$ and $V_{\text{down}}=\tilde{V}_{\text{down}} t_s$ as the maximum vertical ascending and descending displacements, respectively.
Furthermore, recall that $z_{\text{min}}$ and $z_{\text{max}}$ denote the UAV's minimally and maximally allowed altitudes, respectively.
Accordingly, we have
\begin{subequations}\label{Altitude_constraint}
\begin{align}
&z[n+1]-z[n]\leq V_{\text{up}}, \forall n\in \{0\}\cup\mathcal{N},\\
&z[n]-z[n+1]\leq V_{\text{down}}, \forall n\in \{0\}\cup\mathcal{N},\\
&z_{\text{min}}\leq z[n]\leq z_{\text{max}}, \forall n\in \mathcal{N}.
\end{align}
\end{subequations}

We consider that the UAV can adaptively allocate its transmit power over the duration-$T$ communication period. Let $p[n]\ge 0$ denote the transmit power by the UAV at slot $n \in \mathcal{N}$. Suppose that the UAV is subject to a maximum average power $P_\text{ave}$ and a maximum peak power $P_\text{peak}$, where $P_\text{ave} \le P_\text{peak}$ holds in general.
Then we have
\begin{subequations}\label{Power_constraint}
\begin{align}
&\frac{1}{N}\sum_{n\in\mathcal{N}} p[n] \le P_\text{ave}, \label{Power_ave}\\
&0\le p[n] \le P_\text{peak}, \forall n\in\mathcal{N}. \label{Power_peak}
\end{align}
\end{subequations}

Similar as in the quasi-stationary UAV scenario in Section \uppercase\expandafter{\romannumeral2}.A and with CoMP reception, the secrecy rate from the UAV to the $K$ GRs at time slot $n$ (in bps/Hz) is given by
$R^{(\text{\uppercase\expandafter{\romannumeral1}})}\left(\mv{q}[n],z[n],p[n]\right)$ in (\ref{Rate_static_noncolluding}) for Type-I eavesdroppers, or $R^{(\text{\uppercase\expandafter{\romannumeral2}})}\left(\mv{q}[n],z[n],p[n]\right)$ in (\ref{Rate_static_colluding}) for Type-II eavesdroppers.

In the mobile UAV scenario, our objective is to jointly optimize the 3D UAV trajectory $\{\mv{q}[n], z[n]\}$ and the transmit power allocation $\{p[n]\}$ over time, to maximize the average secrecy rate from the UAV to GRs over the whole duration-$T$ communication period (i.e., $\frac{1}{N}\sum_{n\in \mathcal{N}}R^{(i)}\left(\mv{q}[n],z[n],p[n]\right)$, under Type-$i$ eavesdroppers, $i\in\{\text{\uppercase\expandafter{\romannumeral1}},\text{\uppercase\expandafter{\romannumeral2}}\}$), subject to the UAV flight constraints in (\ref{Traj_constraint}) and (\ref{Altitude_constraint}), and the maximum power constraints in (\ref{Power_constraint}). The secrecy rate maximization problem under Type-$i$ eavesdroppers, $i\in\{\text{\uppercase\expandafter{\romannumeral1}},\text{\uppercase\expandafter{\romannumeral2}}\}$, is thus formulated as
\begin{align}
\nonumber(\text{P2-}i):&~\max \limits_{\{\mv{q}[n],z[n],p[n]\}}\frac{1}{N}\sum\limits_{n\in \mathcal{N}}R^{(i)}\left(\mv{q}[n],z[n],p[n]\right)\\
\nonumber ~&\text{s.t.}~ (\ref{Traj_constraint}),~(\ref{Altitude_constraint}),~\text{and}~(\ref{Power_constraint}).
\end{align}
Note that each objective function in problems $(\text{P2-{\uppercase\expandafter{\romannumeral1}}})$ and $(\text{P2-{\uppercase\expandafter{\romannumeral2}}})$ contains a large number of secrecy rate terms that are non-smooth (due to the operator $[\cdot]^+$) and non-concave, with variables $\{\mv{q}[n]\}$, $\{z[n]\}$, and $\{p[n]\}$ coupled.
Therefore, problems $(\text{P2-{\uppercase\expandafter{\romannumeral1}}})$ and $(\text{P2-{\uppercase\expandafter{\romannumeral2}}})$ are non-convex optimization problems that are more difficult to be solved than problems (P1-I) and (P1-II).

In the next two sections, we will address the secrecy rate maximization problems (P1-I) and (P1-II) for the quasi-stationary UAV scenario in Section {\uppercase\expandafter{\romannumeral3}}, and solve problems (P2-I) and (P2-II) for the mobile UAV scenario in Section {\uppercase\expandafter{\romannumeral4}}.

\section{Joint 3D Placement and Transmit Power Control in Quasi-stationary UAV Scenario}
In this section, we obtain the optimal solutions to problems $(\text{P1-{\uppercase\expandafter{\romannumeral1}}})$ and $(\text{P1-{\uppercase\expandafter{\romannumeral2}}})$ in the quasi-stationary UAV scenario under non-colluding and colluding eavesdroppers, respectively.

\subsection{Optimal Solution to Problem $(\text{P1-{\uppercase\expandafter{\romannumeral1}}})$ for Non-colluding Eavesdroppers}
To solve problem $(\text{P1-{\uppercase\expandafter{\romannumeral1}}})$, we first handle the non-smoothness of the objective function.
According to Lemma 1 in \cite{ZhangSecuring2018}, the transmit power control in $(\text{P1-{\uppercase\expandafter{\romannumeral1}}})$ can always lead to a non-negative secrecy rate, since otherwise, we can always set $p=0$ to have a zero secrecy rate. Therefore, problem $(\text{P1-{\uppercase\expandafter{\romannumeral1}}})$ can be equivalently reformulated as
\begin{align}
\nonumber(\text{P1-{\uppercase\expandafter{\romannumeral1}}.1}): &\mathop {\max }\limits_{{\mv{q}},z,p}  {\bar R^{(\text{\uppercase\expandafter{\romannumeral1}})}\left(\mv{q},z,p\right)}\\
\nonumber ~\text{s.t.}~ &z_{\text{min}}\leq z \leq z_{\text{max}}\\
\nonumber&0\le p \le P,
\end{align}
where
\begin{align}\label{R_P1-I.1}
\bar R^{(\text{\uppercase\expandafter{\romannumeral1}})}\left(\mv{q},z,p\right)
=\log_2\left(1+\sum\limits_{k\in\mathcal{K}}g_{bk}\left(\mv{q},z\right)p\right)-\log_2\left(1+\max\limits_{j\in\mathcal{J}}{g_{ej}\left(\mv{q},z\right)p}\right).
\end{align}

Next, we focus on solving problem $(\text{P1-{\uppercase\expandafter{\romannumeral1}}.1})$ that is still non-convex. Towards this end, we first optimize the UAV's altitude $z$ and the transmit power $p$ under any given horizontal location $\mv{q}$, and then find the optimal $\mv{q}$ via a 2D search.

In the following, we only need to focus on optimizing $z$ and $p$ under given $\mv{q}$, for which the optimization problem is simplified as
\begin{align}
\nonumber(\text{P1-{\uppercase\expandafter{\romannumeral1}}.2}):&\mathop {\max }\limits_{p,z} ~{\bar R^{(\text{\uppercase\expandafter{\romannumeral1}})}\left(\mv{q},z,p\right)}\\
\nonumber ~\text{s.t.}~ &z_{\text{min}}\leq z \leq z_{\text{max}}\\
\nonumber&0\le p \le P.
\end{align}
For problem (P1-I.2), it is evident that the optimal transmit power solution of $p$ can either be $0$ or $P$. This is due to the fact that under any given $z$, if the effective legitimate communication link is no weaker than the effective eavesdropping link, i.e., $\sum\limits_{k\in\mathcal{K}}g_{bk}\left(\mv{q},z\right) \ge \max\limits_{j\in\mathcal{J}}g_{ej}\left(\mv{q},z\right)$, then the objective function $\bar R^{(\text{\uppercase\expandafter{\romannumeral1}})}\left(\mv{q},z,p\right)$ in (\ref{R_P1-I.1}) is concave and monotonically non-decreasing with respect to the transmit power $p \ge 0$; while if $\sum\limits_{k\in\mathcal{K}}g_{bk}\left(\mv{q},z\right) < \max\limits_{j\in\mathcal{J}}g_{ej}\left(\mv{q},z\right)$, $\bar R^{(\text{\uppercase\expandafter{\romannumeral1}})}\left(\mv{q},z,p\right)$ in (\ref{R_P1-I.1}) is a convex and monotonically decreasing function of $p \ge 0$.
Furthermore, notice that under $p = 0$, the achieved secrecy rate is always zero, regardless of the UAV's altitude $z$. Therefore, we can solve problem (P1-I.2) by first optimizing $z$ under $p = P$, and then comparing the obtained maximum secrecy rate versus zero.

Now, we consider $p = P$, under which problem $(\text{P1-{\uppercase\expandafter{\romannumeral1}}.2})$ is simplified as
\begin{align}
\nonumber(\text{P1-{\uppercase\expandafter{\romannumeral1}}.3}):&\mathop {\max }\limits_{z} ~{\bar R^{(\text{\uppercase\expandafter{\romannumeral1}})}\left(\mv{q},z,P\right)}\\
\nonumber ~\text{s.t.}&~ z_{\text{min}}\leq z \leq z_{\text{max}}.
\end{align}

We then have the following lemma.
\begin{lemma}\label{Lemma_z_fix_location}
The function ${\bar R^{(\text{\uppercase\expandafter{\romannumeral1}})}\left(\mv{q},z,P\right)}$ is monotonically decreasing, or first increasing and then decreasing, with respect to $z\in [0,+\infty)$.
\begin{IEEEproof}
This lemma can be easily verified by checking the first-derivative of ${\bar R^{(\text{\uppercase\expandafter{\romannumeral1}})}\left(\mv{q},z,P\right)}$ with respect to $z\in [0,+\infty)$. Therefore, the details are omitted for brevity.
\end{IEEEproof}
\end{lemma}
\vspace{3ex}

Based on Lemma \ref{Lemma_z_fix_location}, it is clear that the optimal UAV altitude $z^*$ to problem $(\text{P1-{\uppercase\expandafter{\romannumeral1}}.3})$ is unique, which can be obtained by using a bisection method over $z_{\text{min}}\leq z \leq z_{\text{max}}$.
Therefore, by comparing ${\bar R^{(\text{\uppercase\expandafter{\romannumeral1}})}\left(\mv{q},z^*,P\right)}$ with zero, we have the optimal solutions of $p^\star$ and $z^\star$ to problem (P1-I.2) as follows.
\begin{equation}\label{P1.3_solution}
\begin{gathered}
p^\star= \begin{cases}
  P,&\text{if}~~{\bar R^{(\text{\uppercase\expandafter{\romannumeral1}})}\left(\mv{q},z^*,P\right)}>0,\\
  0,&\text{if}~~{\bar R^{(\text{\uppercase\expandafter{\romannumeral1}})}\left(\mv{q},z^*,P\right)} \le 0,
\end{cases}~~~~~~~~
z^\star\begin{cases}
  =z^*,&\text{if}~~{\bar R^{(\text{\uppercase\expandafter{\romannumeral1}})}\left(\mv{q},z^*,P\right)}>0,\\
  \in[z_{\text{min}},z_{\text{max}}],&\text{if}~~{\bar R^{(\text{\uppercase\expandafter{\romannumeral1}})}\left(\mv{q},z^*,P\right)} \le 0.
\end{cases}
\end{gathered}
\end{equation}
As a result, problem (P1-I.2) is solved. By combining the solution in (\ref{P1.3_solution}) together with the 2D search over \mv{q}, the optimal solution to problem (P1-I.1) or equivalently (P1-I) is finally obtained.

\begin{remark}\label{Remark_P1.3_solution}
First, we discuss the optimal solution to problem (P1-I.2) under given $\mv{q}$ to gain more insights. Notice that under given $\mv{q}$, if there exists one legitimate GR that is located closer to the UAV than all the non-colluding eavesdroppers, i.e., $\min\limits_{k\in\mathcal{K}}{d_{bk}\left(\mv{q},0\right)} \le \min\limits_{j\in\mathcal{J}}d_{ej}\left(\mv{q},0\right),$\footnote{Notice that with $\min\limits_{k\in\mathcal{K}}{d_{bk}\left(\mv{q},0\right)} \le \min\limits_{j\in\mathcal{J}} d_{ej}\left(\mv{q},0\right)$, we have $\min\limits_{k\in\mathcal{K}}{d_{bk}\left(\mv{q},z\right)} \le \min\limits_{j\in\mathcal{J}} d_{ej}\left(\mv{q},z\right)$, and accordingly $\sum\limits_{k\in\mathcal{K}}g_{bk}\left(\mv{q},z\right) > \max\limits_{j\in\mathcal{J}}g_{ej}\left(\mv{q},z\right)$.} (or equivalently, there is at least one legitimate channel that is no weaker than the eavesdropping channels), then it can be shown via checking the first-derivative of ${\bar R^{(\text{\uppercase\expandafter{\romannumeral1}})}\left(\mv{q},z,P\right)}$ with respect to $z\in [0,+\infty)$ that the optimal altitude $z$ to problem (P1-I.2) is $z^\star = z_{\text{min}}$.
In other words, the UAV should stay at the lowest altitude to enjoy the strongest legitimate channel gains.
Otherwise, if the UAV is located closer to one or more eavesdroppers than all legitimate GRs (but $\sum\limits_{k\in\mathcal{K}}g_{bk}\left(\mv{q},z\right) \ge \max\limits_{j\in\mathcal{J}}g_{ej}\left(\mv{q},z\right)$ still holds), then it can be shown that at the optimality of problem (P1-I.2), the UAV altitude may vary between the minimum and maximum values, depending on its horizontal location $\mv{q}$ and the GRs' and eavesdroppers' distributions.
\end{remark}

\begin{remark}
Next, we compare the optimal solution to problem (P1-I) in the special case with $K=1$ GR (without CoMP) versus that in the case with $K>1$ GRs (with CoMP of our interest). First, consider the special case with $K=1$ GR to gain further insights. In this case, it can be shown that at the optimal solution to problem (P1-I.2) under any given $\mv{q}$, the UAV can always stay at the lowest altitude $z^\star = z_{\text{min}}$, even when $g_{b1}\left(\mv{q},z\right) < \max\limits_{j\in\mathcal{J}}g_{ej}\left(\mv{q},z\right)$ (with zero secrecy rate achieved). Therefore, at the optimal solution to problem (P1-I) with $\mv{q}$ optimized, the UAV should also stay at the lowest altitude, with the optimized horizontal location closer to the GR than all eavesdroppers to achieve a positive secrecy rate.
By contrast, in our considered CoMP case with $K>1$ GRs, the UAV may stay at an optimized altitude higher than $z_{\text{min}}$ with the optimized horizontal location closer to some eavesdroppers than GRs, but still achieving a positive secrecy rate. This indicates that controlling the UAV's altitude is not beneficial for enhancing secrecy UAV communication performance if CoMP reception is not employed at GRs, but is very significant if CoMP reception is considered.
\end{remark}

\subsection{Optimal Solution to Problem $(\text{P1-{\uppercase\expandafter{\romannumeral2}}})$ for Colluding Eavesdroppers}
Next, we consider problem (P1-II) for colluding eavesdroppers. Similarly as for $(\text{P1-{\uppercase\expandafter{\romannumeral1}}})$, we omit the $[\cdot]^+$ operator in the objective function of $(\text{P1-{\uppercase\expandafter{\romannumeral2}}})$, and re-formulate $(\text{P1-{\uppercase\expandafter{\romannumeral2}}})$ as the following equivalent problem $(\text{P1-{\uppercase\expandafter{\romannumeral2}}.1})$.
\begin{align}
\nonumber(\text{P1-{\uppercase\expandafter{\romannumeral2}}.1}): &\mathop {\max }\limits_{ {\mv{q}},z,p }  {\bar R^{(\text{\uppercase\expandafter{\romannumeral2}})}\left(\mv{q},z,p\right)}\\
\nonumber ~\text{s.t.}~ &z_{\text{min}}\leq z \leq z_{\text{max}}\\
\nonumber&0\le p \le P,
\end{align}
where
\begin{align}\label{R_P1-II.1}
\bar R^{(\text{\uppercase\expandafter{\romannumeral2}})}\left(\mv{q},z,p\right)
=\log_2\left(1+\sum\limits_{k\in\mathcal{K}}g_{bk}\left(\mv{q},z\right)p\right)-\log_2\left(1+\sum\limits_{j\in\mathcal{J}}{g_{ej}\left(\mv{q},z\right)p}\right).
\end{align}
To solve problem $(\text{P1-{\uppercase\expandafter{\romannumeral2}}.1})$, we first optimize the UAV's altitude $z$ and transmit power $p$ under any given horizontal location $\mv{q}$, and then find the optimal $\mv{q}$ via a 2D search.

Now, we optimize $p$ and $z$ under given $\mv{q}$, for which the optimization problem is expressed as
\begin{align}
\nonumber(\text{P1-{\uppercase\expandafter{\romannumeral2}}.2}):&\mathop {\max }\limits_{p,z} ~{\bar R^{(\text{\uppercase\expandafter{\romannumeral2}})}\left(\mv{q},z,p\right)}\\
\nonumber ~\text{s.t.}~ &z_{\text{min}}\leq z \leq z_{\text{max}}\\
\nonumber&0\le p \le P.
\end{align}
Similar to $(\text{P1-{\uppercase\expandafter{\romannumeral1}}.2})$, the optimal transmit power solution to problem (P1-II.2) is either $0$ or full power $P$. Therefore, to solve (P1-II.2), we only need to optimize $z$ for problem (P1-II.2) under $p = P$, and then compare the obtained secrecy rate versus zero. Under $p = P$,  $(\text{P1-{\uppercase\expandafter{\romannumeral2}}.2})$ is simplified as
\begin{align}
\nonumber(\text{P1-{\uppercase\expandafter{\romannumeral2}}.3}):&\mathop {\max }\limits_{z} {\bar R^{(\text{\uppercase\expandafter{\romannumeral2}})}\left(\mv{q},z,P\right)}\\
\nonumber ~\text{s.t.}~ &z_{\text{min}}\leq z \leq z_{\text{max}}.
\end{align}

It is observed from (\ref{R_P1-II.1}) that the objective function is a complicated function with respect to the variable $z$, which can only be solved numerically.
Note that $\bar R^{(\text{\uppercase\expandafter{\romannumeral2}})}\left(\mv{q},z,P\right)$ is a bounded continuous differentiable function with respect to $z\in[z_{\text{min}},z_{\text{max}}]$. Therefore, we can check the first derivative of $\bar R^{(\text{\uppercase\expandafter{\romannumeral2}})}\left(\mv{q},z,P\right)$ and obtain all the real solutions to $\bar R^{(\text{\uppercase\expandafter{\romannumeral2}})}\left(\mv{q},z,P\right) = 0$. By comparing the corresponding objective values under these solutions and boundary points (i.e., $z_{\text{min}}$ and $z_{\text{max}}$), we can get the optimal value of $z$ to problem $(\text{P1-{\uppercase\expandafter{\romannumeral2}}.3})$, denoted by $z^{**}$.

Then, by comparing the obtained ${\bar R^{(\text{\uppercase\expandafter{\romannumeral2}})}\left(\mv{q},z^{**},P\right)}$ with zero, we have the optimal solution of $p^{\star\star}$ and $z^{\star\star}$ to problem (P1-II.2) as
\begin{equation}\label{P1.II.3_solution}
\begin{gathered}
p^{\star\star}= \begin{cases}
  P,&\text{if}~~{\bar R^{(\text{\uppercase\expandafter{\romannumeral2}})}\left(\mv{q},z^{**},P\right)}>0,\\
  0,&\text{if}~~{\bar R^{(\text{\uppercase\expandafter{\romannumeral2}})}\left(\mv{q},z^{**},P\right)} \le 0,
\end{cases}~~~~~~
z^{\star\star}\begin{cases}
  =z^{**},&\text{if}~~{\bar R^{(\text{\uppercase\expandafter{\romannumeral2}})}\left(\mv{q},z^{**},P\right)}>0,\\
  \in[z_{\text{min}},z_{\text{max}}],&\text{if}~~{\bar R^{(\text{\uppercase\expandafter{\romannumeral2}})}\left(\mv{q},z^{**},P\right)} \le 0.
\end{cases}
\end{gathered}
\end{equation}
Finally, by combining $p^{\star\star}$ and $z^{\star\star}$ together with the 2D search for $\mv{q}$, the optimal solution to problem (P1-II) is finally obtained.
Notice that as the secrecy rate function ${R^{(\text{\uppercase\expandafter{\romannumeral2}})}\left(\mv{q},z,p\right)}$ under colluding eavesdroppers is generally more complicated than ${ R^{(\text{\uppercase\expandafter{\romannumeral1}})}\left(\mv{q},z,p\right)}$ under non-colluding eavesdroppers, it is difficult to analytically analyze the optimal solution to problem (P1-II). Therefore, we will compare the optimal solutions under colluding and non-colluding eavesdroppers in numerical results in Section {\uppercase\expandafter{\romannumeral5}} later.

\section{Joint 3D Trajectory and Transmit Power Allocation in Mobile UAV Scenario}
In this section, we propose efficient algorithms to solve problems $(\text{P2-{\uppercase\expandafter{\romannumeral1}}})$ and $(\text{P2-{\uppercase\expandafter{\romannumeral2}}})$ in the mobile UAV scenario for both non-colluding and colluding eavesdroppers, respectively.

\subsection{Proposed Solution to $(\text{P2-{\uppercase\expandafter{\romannumeral1}}})$ for Non-colluding Eavesdroppers}
First, we handle the non-smoothness of the objective function of problem $(\text{P2-{\uppercase\expandafter{\romannumeral1}}})$.
Towards this end, we omit the $[\cdot]^+$ operator in the objective function of $(\text{P2-{\uppercase\expandafter{\romannumeral1}}})$ similarly as in the quasi-stationary UAV scenario in Section \uppercase\expandafter{\romannumeral3}.A, and re-express $(\text{P2-{\uppercase\expandafter{\romannumeral1}}})$ as the following equivalent problem $(\text{P2-{\uppercase\expandafter{\romannumeral1}}.1})$.
\begin{align}
\nonumber(\text{P2-{\uppercase\expandafter{\romannumeral1}}.1}): &\mathop {\max }\limits_{\{ {\mv{q}}[n],z[n],p[n]\} } \frac{1}{N}\sum\limits_{n\in \mathcal{N}} {\bar R^{(\text{\uppercase\expandafter{\romannumeral1}})}}\left(\mv{q}[n],z[n],p[n]\right)\\
\nonumber ~&\text{s.t.}~ (\ref{Traj_constraint}),~(\ref{Altitude_constraint}),~\text{and}~(\ref{Power_constraint}),
\end{align}
where $\bar R^{(\text{\uppercase\expandafter{\romannumeral1}})}\left(\mv{q}[n],z[n],p[n]\right)$ is defined in (\ref{R_P1-I.1}).

Next, we focus on solving problem $(\text{P2-{\uppercase\expandafter{\romannumeral1}}.1})$, which, however, is still non-convex. To tackle this issue, we use the alternating optimization method to optimize the transmit power allocation $\{p[n]\}$ and UAV trajectory $\{\mv{q}[n], z[n]\}$ in an iterative manner, by considering the other to be given.

\subsubsection{Transmit Power Allocation Optimization}
First, we optimize the UAV's transmit power $\{p[n]\}$ under given UAV trajectory $\{\mv{q}[n], z[n]\}$, for which the optimization problem is expressed as
\begin{align}\label{subproblem_power}
\nonumber(\text{P2-{\uppercase\expandafter{\romannumeral1}}.2}):\mathop {\max }\limits_{\{p[n]\} }\frac{1}{N}\sum\limits_{n\in \mathcal{N}} {\bar R^{(\text{\uppercase\expandafter{\romannumeral1}})}}\left(\mv{q}[n],z[n],p[n]\right), ~\text{s.t.}&~ (\ref{Power_constraint}).
\end{align}
Notice that under given $\{\mv{q}[n]\}$ and $\{z[n]\}$, $\bar R^{(\text{\uppercase\expandafter{\romannumeral1}})}\left(\mv{q}[n],z[n],p[n]\right)$ can be re-expressed as follows for notational convenience
\begin{align}
\bar R^{(\text{\uppercase\expandafter{\romannumeral1}})}\left(\mv{q}[n],z[n],p[n]\right)= {\log _2} \left( {1 + {a_n^{(\text{\uppercase\expandafter{\romannumeral1}})}}p[n]} \right) - {\log _2} \left( {1 + {b_n^{(\text{\uppercase\expandafter{\romannumeral1}})}}p[n]} \right),
\end{align}
where ${a_n^{(\text{\uppercase\expandafter{\romannumeral1}})}}$ and $b_n^{(\text{\uppercase\expandafter{\romannumeral1}})}$ are constants given as
$a_n^{(\text{\uppercase\expandafter{\romannumeral1}})}=\sum\limits_{k\in\mathcal{K}}g_{bk}\left(\mv{q}[n],z[n]\right),$
$b_n^{(\text{\uppercase\expandafter{\romannumeral1}})}=\max\limits_{j\in\mathcal{J}}g_{ej}\left(\mv{q}[n],z[n]\right).$

It is evident that for problem $(\text{P2-{\uppercase\expandafter{\romannumeral1}}.2})$, under any time slot ${n\in \mathcal{N}}$, if the effective legitimate communication channel is stronger than the effective eavesdropping channel, i.e., $a_n^{(\text{\uppercase\expandafter{\romannumeral1}})} > b_n^{(\text{\uppercase\expandafter{\romannumeral1}})}$, then the rate function $\bar R^{(\text{\uppercase\expandafter{\romannumeral1}})}\left(\mv{q}[n],z[n],p[n]\right)$ is concave and monotonically increasing with respect to the transmit power $p[n] \ge 0$; otherwise, it can be easily shown that the maximum of $\bar R^{(\text{\uppercase\expandafter{\romannumeral1}})}\left(\mv{q}[n],z[n],p[n]\right)$ is zero, which is attained at $p[n] = 0$. Therefore, we only need to consider the transmit power allocation over a subset $\overline{\mathcal N}$ of time slots, with
$\overline{\mathcal N}  = \{n | a_n^{(\text{\uppercase\expandafter{\romannumeral1}})} > b_n^{(\text{\uppercase\expandafter{\romannumeral1}})}, n\in \mathcal N\}$. In this case, problem $(\text{P2-{\uppercase\expandafter{\romannumeral1}}.2})$ is equivalently re-expressed as
\begin{subequations}
\begin{align}
\nonumber (\text{P2-{\uppercase\expandafter{\romannumeral1}}.3}):&\mathop {\max }\limits_{\{p[n]\} }\frac{1}{N}\sum\limits_{n\in \overline{\mathcal N}} {\bar R^{(\text{\uppercase\expandafter{\romannumeral1}})}}\left(\mv{q}[n],z[n],p[n]\right)\\
 \text{s.t.}~ &\frac{1}{N}\sum_{n\in\overline{\mathcal{N}}} p[n] \le P_\text{ave}\label{P41}\\
 &0\le p[n] \le P_\text{peak}, \forall {n\in \overline{\mathcal N}}. \label{P42}
\end{align}
\end{subequations}

The reformulated problem $(\text{P2-{\uppercase\expandafter{\romannumeral1}}.3})$ is convex and satisfies the Slater's conditions \cite{boyd2004convex}. Therefore, this problem can be optimally solved by the Karush-kuhn-Tucker (KKT) conditions \cite{boyd2004convex}.
Let $\upsilon$ denote the Lagrange multiplier associated with constraint (\ref{P41}), and $\underline\lambda_n$ and $\overline\lambda_n, {n\in \overline{\mathcal N}}$ denote the Lagrange multipliers associated with $p[n] \ge 0$ and  $p[n] \le P_\text{peak}$, respectively. Suppose that the optimal primal and dual solutions to problem $(\text{P2-{\uppercase\expandafter{\romannumeral1}}.3})$ are given by $\{p^*[n]\}$, $\upsilon^*$, $\{\underline\lambda^*_n\}$, and $\{\overline\lambda_n^*\}$, respectively. Then based on the KKT conditions, they should satisfy the following sufficient and necessary conditions.
\begin{subequations}\label{KKT}
\begin{align}
\label{KKT1_1}&\upsilon^*\ge 0,\\
\label{KKT1_1_1}&\upsilon^*\left(\frac{1}{N}\sum_{n\in \overline{\mathcal N}} p^*[n] - P_\text{ave}\right)=0,\\
\label{KKT1_2}&0\le p^*[n] \le P_\text{peak}, \forall n\in\overline{\mathcal N},\\
\label{KKT1_3}&{\underline\lambda_n^*} \ge 0,~{\overline\lambda_n^*} \ge 0,~\forall n\in\overline{\mathcal N},\\
\label{KKT1_4}&{\underline\lambda_n^*} p^*[n] =  0,~\forall n\in\overline{\mathcal N},\\
\label{KKT1_4_2}&{\overline\lambda_n^*} (p^*[n]-P_\text{peak}) =  0,~\forall n\in\overline{\mathcal N},\\
\label{KKT1_5}&\frac{1}{{\ln 2}}\left( {\frac{{{a_n^{(\text{\uppercase\expandafter{\romannumeral1}})}}}}{{1 + {a_n^{(\text{\uppercase\expandafter{\romannumeral1}})}}p^*[n]}} - \frac{{{b_n^{(\text{\uppercase\expandafter{\romannumeral1}})}}}}{{1 + {b_n^{(\text{\uppercase\expandafter{\romannumeral1}})}}p^*[n]}}} \right) + {\underline\lambda _n^*}- {\overline\lambda_n^*}= \upsilon^*,\forall n\in\overline{\mathcal N}.
\end{align}
\end{subequations}
Based on the above KKT conditions in (\ref{KKT}) and after some manipulations, the optimal solution to problem $(\text{P2-{\uppercase\expandafter{\romannumeral1}}.3})$ is given as
\begin{align} \label{KKT_power}
p^*[n] = \min (P_\text{peak},~\tilde p^*[n]),~\forall n\in\overline{\mathcal N},
\end{align}
where $\tilde p^*[n] = \bigg[ {\frac{{ - {a_n^{(\text{\uppercase\expandafter{\romannumeral1}})}} - {b_n^{(\text{\uppercase\expandafter{\romannumeral1}})}} + \sqrt {{{({a_n^{(\text{\uppercase\expandafter{\romannumeral1}})}} - {b_n^{(\text{\uppercase\expandafter{\romannumeral1}})}})}^2} + 4{a_n^{(\text{\uppercase\expandafter{\romannumeral1}})}}{b_n^{(\text{\uppercase\expandafter{\romannumeral1}})}}(\frac{{{a_n^{(\text{\uppercase\expandafter{\romannumeral1}})}} - {b_n^{(\text{\uppercase\expandafter{\romannumeral1}})}}}}{{\upsilon^* \ln 2}})} }}{{2{{a_n^{(\text{\uppercase\expandafter{\romannumeral1}})}}_n}{b_n^{(\text{\uppercase\expandafter{\romannumeral1}})}}}}} \bigg]^+$.
By combining (\ref{KKT_power}) with the fact that the optimal $p[n]$ should be zero for any $n \in {\mathcal N} \setminus \overline{\mathcal N}$, it thus follows that the optimal solution to problem $(\text{P2-{\uppercase\expandafter{\romannumeral1}}.2})$ is
\begin{align}
p^\star[n] = \left\{ \begin{array}{l} p^*[n],~\forall  n\in \overline{\mathcal N},\\
0,~~~~~~\forall n \in {\mathcal N} \setminus  \overline{\mathcal N}.
\end{array} \right.
\end{align}
Notice that $p^*[n]$'s in (\ref{KKT_power}) only depends on the optimal Lagrange multiplier $\upsilon^*$, which can be obtained via a bisection search based on $\frac{1}{N}\sum_{n\in \overline{\mathcal N}} p^*[n] - P_\text{ave}=0$.\footnote{Notice that $\upsilon^*$ can become zero if $\sum_{n\in \overline{\mathcal N}} P_\text{peak} < N P_\text{ave}$ (or equivalently, the equality $\frac{1}{N}\sum_{n\in \overline{\mathcal N}} p^*[n] - P_\text{ave}=0$ cannot be met). To avoid this case, we need to check whether $\sum_{n\in \overline{\mathcal N}} P_\text{peak} < N P_\text{ave}$ holds before the bisection search.}
Therefore, the transmit power allocation problem $(\text{P2-{\uppercase\expandafter{\romannumeral1}}.2})$ is finally solved optimally.

\subsubsection{Trajectory Optimization}
Next, we optimize the UAV trajectory $\{\mv{q}[n], z[n]\}$ under any given transmit power allocation $\{p[n]\}$, for which the optimization problem is expressed as
\begin{align}
\nonumber(\text{P2-{\uppercase\expandafter{\romannumeral1}}.4}):&\mathop {\max }\limits_{\{ {\mv{q}}[n],z[n]\} } \frac{1}{N}\sum\limits_{n\in \mathcal{N}} {\bar R^{(\text{\uppercase\expandafter{\romannumeral1}})}}\left(\mv{q}[n],z[n],p[n]\right)\\
\nonumber ~\text{s.t.}&~ (\ref{Traj_constraint})~\text{and}~(\ref{Altitude_constraint}).
\end{align}

By introducing a set of auxiliary variables $\{r[n]\}$, problem $(\text{P2-{\uppercase\expandafter{\romannumeral1}}.4})$ is equivalently reformulated as the following problem:
\begin{align}
\nonumber(\text{P2-{\uppercase\expandafter{\romannumeral1}}.5}):&\mathop {\max }\limits_{\{ {\mv{q}}[n],z[n],r[n]\} }   \frac{1}{N}\sum\limits_{n\in \mathcal{N}}r[n]\\
\nonumber\text{s.t.}~&\log_2\!\left(\!1\!+\!\gamma_b\!\left(\mv{q}[n],z[n],p[n]\right)\!\right)\!-\!\log_2\!\left(\!1\!+\!\gamma_{ej}\!\left(\mv{q}[n],z[n],p[n]\right)\!\right)\ge r[n], \forall {j\in\mathcal{J},n\in \mathcal{N}}\\
\nonumber&(\ref{Traj_constraint})~\text{and}~(\ref{Altitude_constraint}).
\end{align}

By further introducing auxiliary variables $\{{\zeta}_k[n]\}$ and $\{{\eta}_j[n]\}$, problem $(\text{P2-{\uppercase\expandafter{\romannumeral1}}.5})$ is equivalently re-expressed as
\begin{subequations}
\begin{align}
(\text{P2-{\uppercase\expandafter{\romannumeral1}}.6}):&\max\limits_{\{\mv{q}[n],z[n],r[n],{\zeta}_k[n],{\eta}_j[n]\}}\frac{1}{N}\sum\limits_{n\in \mathcal{N}}r[n]\nonumber\\
\text{s.t.}~&\hat{R}_j^{(\text{\uppercase\expandafter{\romannumeral1}})}\left({\zeta}_k[n],{\eta}_j[n],p[n]\right) \ge r[n], \forall {j\in\mathcal{J},n\in \mathcal{N}}\label{Nonconvex2}\\
&{\zeta}_k[n]\geq \left(||\mv{q}[n]-\mv{w}_{bk}||^2+z^2[n]\right)^{\frac{\alpha}{2}},\forall k\in \mathcal{K}, n\in \mathcal{N}\label{P23_1}\\
&{\eta}_j[n]\leq \left(||\mv{q}[n]-\mv{w}_{ej}||^2+z^2[n]\right)^{\frac{\alpha}{2}},\forall j\in \mathcal{J}, n\in \mathcal{N}\label{Nonconvex1}\\
&(\ref{Traj_constraint}),~(\ref{Altitude_constraint}),\nonumber
\end{align}
\end{subequations}
where $\hat{R}_j^{(\text{\uppercase\expandafter{\romannumeral1}})}\left({\zeta}_k[n],{\eta}_j[n],p[n]\right)={\log _2}\left( {1 + \sum\limits_{k \in {\cal K}} {\frac{{{\beta _0}p[n]}}{{{\zeta _k}[n]}}} } \right) - {\log _2}\bigg( {1 +  {\frac{{{\beta _0}p[n]}}{{{\eta_j}[n]}}} } \bigg)$.
Notice that each function $\hat{R}_j^{(\text{\uppercase\expandafter{\romannumeral1}})}\left({\zeta}_k[n],{\eta}_j[n],p[n]\right)$ in the left-hand-side (LHS) terms in constraint (\ref{Nonconvex2}) and the right-hand-side (RHS) terms in constraint (\ref{Nonconvex1}) are convex with respect to $\{\mv{q}[n], z[n]\}$, thus making problem $(\text{P2-{\uppercase\expandafter{\romannumeral1}}.6})$ non-convex.
To tackle the non-convexity issue, we apply the SCA technique to obtain a converged solution in an iterative manner.
At each iteration $m\geq1$, suppose that the local trajectory point is given as $\{\mv q^{(m)}[n], z^{(m)}[n]\}$.
Then, we have the lower bounds for the function $\hat{R}_j^{(\text{\uppercase\expandafter{\romannumeral1}})}\left({\zeta}_k[n],{\eta}_j[n],p[n]\right)$ and the RHS terms of (\ref{Nonconvex1}) as follows based on the first-order Taylor expansion.
\begin{align}
\nonumber&\hat R_j^{(\text{\uppercase\expandafter{\romannumeral1}})}\left({\zeta}_k[n],{\eta}_j[n],p[n]\right)\ge {{\hat R_j}^{(\text{\uppercase\expandafter{\romannumeral1}}m)}}\left({\zeta}_k[n],{\eta}_j[n],p[n]\right)
\triangleq {\log _2}\bigg( {1 + \sum\limits_{k \in {\cal K}} {\frac{{{\beta _0}p[n]}}{{\zeta_k^{(m)}[n]}}} } \bigg)\\
&- {\log _2}\bigg( {1 + {\frac{{{\beta _0}p[n]}}{{{\eta_j}[n]}}} } \bigg)
-\frac{1}{\ln2} {{\bigg( {1 + \sum\limits_{k \in {\cal K}} {\frac{{{\beta _0}p[n]}}{{\zeta _k^{(m)}[n]}}} } \bigg)}^{-1}}\sum\limits_{k \in {\cal K}} {\bigg({\frac{{{\beta _0}p[n]}}{{{{\zeta _k^{(m)2}[n]}}}}\left( {{\zeta _k}[n] - \zeta_k^{(m)}[n]} \right)} \bigg)}\label{convex2},
\end{align}
\begin{align}\label{convex1}
&\left(||\mv{q}[n]-\mv{w}_{ej}||^2+z^{2}[n]\right)^{\alpha/2}
\geq\alpha\Big[{z^{(m)}}[n](z[n]- {z^{(m)}}[n]) + ({{\mv{q}}^{(m)}}[n] - {{\mv{w}}_{ej}})({\mv{q}}[n] - {{\mv{q}}^{(m)}}[n])\Big]
\nonumber\\
&\times\left(||\mv{q}^{(m)}[n]-\mv{w}_{ej}||^2+z^{(m)2}[n]\right)^{(\alpha-2)/2}
+\left(||\mv{q}^{(m)}[n]-\mv{w}_{ej}||^2+z^{(m)2}[n]\right)^{\alpha/2}\triangleq {E_{ej}^{\text{lb}}}^{(\text{\uppercase\expandafter{\romannumeral1}})}[n].
\end{align}
Replacing $\hat{R}_j^{(\text{\uppercase\expandafter{\romannumeral1}})}\left(\mv{q}[n],z[n],p[n]\right)$ and the RHS terms in (\ref{Nonconvex1}) as ${{\hat R_j}^{(\text{\uppercase\expandafter{\romannumeral1}}m)}}$ and ${E_{ej}^{\text{lb}}}^{(\text{\uppercase\expandafter{\romannumeral1}})}[n]$, respectively, problem $(\text{P2-{\uppercase\expandafter{\romannumeral1}}.6})$ is approximately expressed as the following convex optimization problem that can be efficiently solved by CVX \cite{grant2014cvx}.
\begin{align}
\nonumber(\text{P2-{\uppercase\expandafter{\romannumeral1}}.7.}m):&\max\limits_{\{ {\mv{q}}[n],z[n],r[n], {\zeta_k}[n],{\eta_j}[n]\} }  \frac{1}{N}\sum\limits_{n\in \mathcal{N}}r[n]\\
\nonumber\text{s.t.}~&{{{\hat R_j}^{(\text{\uppercase\expandafter{\romannumeral1}}m)}}}\left({\zeta}_k[n],{\eta}_j[n],p[n]\right) \ge r[n], \forall {j\in\mathcal{J}}\\
\nonumber&{\eta}_j[n]\leq {E_{ej}^{\text{lb}}}^{(\text{\uppercase\expandafter{\romannumeral1}})}[n],\forall j\in \mathcal{J}, n\in \mathcal{N}\\
\nonumber&(\ref{Traj_constraint}),~(\ref{Altitude_constraint}),~\text{and}~(\ref{P23_1}).
\end{align}

Therefore, at iteration $(m+1)$, we update the UAV trajectory point $\{\mv{q}^{(m+1)}[n], z^{(m+1)}[n]\}$ as the optimal solution to the approximate problem (P2-{{\uppercase\expandafter{\romannumeral1}}}.7.$m$), under the local trajectory point $\{\mv{q}^{(m)}[n], z^{(m)}[n]\}$ in the previous iteration $m$.
As the iteration converges, we can obtain an efficient solution to problem $(\text{P2-{\uppercase\expandafter{\romannumeral1}}.5})$.

To sum up, we solve for the transmit power $\{p[n]\}$ and the trajectory $\{\mv{q}[n], z[n]\}$ in an iterative manner above, and accordingly, we obtain an efficient solution to problem $(\text{P2-{\uppercase\expandafter{\romannumeral1}}.1})$ or (P2-I).
As the objective value of problem $(\text{P2-{\uppercase\expandafter{\romannumeral1}}.1})$ is monotonically non-decreasing after each iteration and the objective value of problem $(\text{P2-{\uppercase\expandafter{\romannumeral1}}.1})$ is finite, the proposed alternating optimization based approach is guaranteed to converge \cite{bertsekas1999nonlinear}. In Section \ref{nummerical}, we will conduct simulations to show the effectiveness of the proposed algorithm.

\begin{remark}\label{remark1}
It should be noticed that the performance of our proposed alternating-optimization-based approach critically depends on the initial point for iteration. In this paper, we consider the following fly-hover-fly trajectory as the initial point. In this design, for the horizontal trajectory, the UAV first flies straightly at the maximum speed from the initial location to the top of one GR, then hovers with the maximum duration, and finally flies straightly at the maximum speed to the final location.
We choose the hovering location as the point above the GR at the most central point among these GRs.
For the vertical trajectory, if the altitude of the initial location is different from the final location, the UAV first flies at the maximum speed to reach the altitude of the final location, then stays at this altitude in the rest of the mission duration.
\end{remark}

\subsection{Proposed Solution to $(\text{P2-{\uppercase\expandafter{\romannumeral2}}})$ for Colluding Eavesdroppers}
In this subsection, we propose an efficient solution to problem $(\text{P2-{\uppercase\expandafter{\romannumeral2}}})$ for colluding eavesdroppers.
First, we handle the non-smoothness of the objective function of problem $(\text{P2-{\uppercase\expandafter{\romannumeral2}}})$.
As explained before, we omit the $[\cdot]^+$ operator in the objective function of $(\text{P2-{\uppercase\expandafter{\romannumeral2}}})$, and equivalently re-express $(\text{P2-{\uppercase\expandafter{\romannumeral2}}})$ as the following problem:
\begin{align}
\nonumber(\text{P2-{\uppercase\expandafter{\romannumeral2}}.1}): &\mathop {\max }\limits_{\{ {\mv{q}}[n],z[n],p[n]\} } \frac{1}{N}\sum\limits_{n\in \mathcal{N}} {\bar R^{(\text{\uppercase\expandafter{\romannumeral2}})}}\left(\mv{q}[n],z[n],p[n]\right)\\
\nonumber ~\text{s.t.}&~ (\ref{Traj_constraint}),~(\ref{Altitude_constraint}),~\text{and}~(\ref{Power_constraint}),
\end{align}
where $\bar R^{(\text{\uppercase\expandafter{\romannumeral2}})}\left(\mv{q}[n],z[n],p[n]\right)$ is defined in (\ref{R_P1-II.1}).

Next, we focus on solving the non-convex problem $(\text{P2-{\uppercase\expandafter{\romannumeral2}}.1})$, by using the alternating optimization method. In particular, we optimize the transmit power allocation $\{p[n]\}$ and UAV trajectory $\{\mv{q}[n], z[n]\}$ in an iterative manner, by considering the other to be given.

\subsubsection{Transmit Power Allocation Optimization}
First, we optimize the UAV's transmit power $\{p[n]\}$ under given UAV trajectory $\{\mv{q}[n], z[n]\}$, for which the optimization problem is expressed as
\begin{align}\label{subproblem_power_C}
\nonumber(\text{P2-{\uppercase\expandafter{\romannumeral2}}.2}):\mathop {\max }\limits_{\{p[n]\} } \frac{1}{N}\sum\limits_{n\in \mathcal{N}} {\bar R^{(\text{\uppercase\expandafter{\romannumeral2}})}}\left(\mv{q}[n],z[n],p[n]\right), ~\text{s.t.}~ (\ref{Power_constraint}).
\end{align}
Notice that under given $\{\mv{q}[n]\}$ and $\{z[n]\}$, $\bar R^{(\text{\uppercase\expandafter{\romannumeral2}})}\left(\mv{q}[n],z[n],p[n]\right)$ can be re-expressed as follows for notational convenience
\begin{align}
\bar R^{(\text{\uppercase\expandafter{\romannumeral2}})}\left(\mv{q}[n],z[n],p[n]\right)= {\log _2}\left( {1 + {a_n^{(\text{\uppercase\expandafter{\romannumeral2}})}}p[n]} \right) - {\log _2}\left( {1 + {b_n^{(\text{\uppercase\expandafter{\romannumeral2}})}}p[n]} \right),
\end{align}
where $a_n^{(\text{\uppercase\expandafter{\romannumeral2}})}=\sum\limits_{k\in\mathcal{K}}g_{bk}\left(\mv{q}[n],z[n]\right)$ and $b_n^{(\text{\uppercase\expandafter{\romannumeral2}})}=\sum\limits_{j\in\mathcal{J}}g_{ej}\left(\mv{q}[n],z[n]\right)$.
Similar to problem $(\text{P2-{\uppercase\expandafter{\romannumeral1}}.2})$, problem $(\text{P2-{\uppercase\expandafter{\romannumeral2}}.2})$ is equivalently re-expressed as
\begin{subequations}
\begin{align}
\nonumber (\text{P2-{\uppercase\expandafter{\romannumeral2}}.3}):&\mathop {\max }\limits_{\{p[n]\} }\frac{1}{N}\sum\limits_{n\in \hat{\mathcal N}} {\bar R^{(\text{\uppercase\expandafter{\romannumeral2}})}}\left(\mv{q}[n],z[n],p[n]\right)\\
 ~\text{s.t.}~ &\frac{1}{N}\sum_{n\in\hat{\mathcal{N}}} p[n] \le P_\text{ave} \label{P41_C}\\
 &0\le p[n] \le P_\text{peak}, \forall {n\in \hat{\mathcal N}}, \label{P42_C}
\end{align}
\end{subequations}
where $\hat{\mathcal N}$ is a subset of time slots, with $\hat{\mathcal N}  = \{n | a_n^{(\text{\uppercase\expandafter{\romannumeral2}})} > b_n^{(\text{\uppercase\expandafter{\romannumeral2}})}, n\in \mathcal N\}$.
Accordingly, the optimal solution to problem $(\text{P2-{\uppercase\expandafter{\romannumeral2}}.3})$ obtained as $\{p^{**}[n]\}$, similarly as in (\ref{KKT_power}), by replacing $\overline{\mathcal N}$, $a_n^{(\text{\uppercase\expandafter{\romannumeral1}})}$, and $b_n^{(\text{\uppercase\expandafter{\romannumeral1}})}$ as $\hat{\mathcal N}$, $a_n^{(\text{\uppercase\expandafter{\romannumeral2}})}$, and $b_n^{(\text{\uppercase\expandafter{\romannumeral2}})}$, respectively.
Therefore, the transmit power allocation problem $(\text{P2-{\uppercase\expandafter{\romannumeral2}}.2})$ is finally solved optimally.

\subsubsection{Trajectory Optimization}
Next, we optimize the UAV trajectory $\{\mv{q}[n], z[n]\}$ under any given transmit power allocation $\{p[n]\}$, for which the optimization problem is expressed as
\begin{align}
\nonumber(\text{P2-{\uppercase\expandafter{\romannumeral2}}.4}):&\mathop {\max }\limits_{\{ {\mv{q}}[n],z[n]\} } \frac{1}{N}\sum\limits_{n\in \mathcal{N}} {\bar R^{(\text{\uppercase\expandafter{\romannumeral2}})}}\left(\mv{q}[n],z[n],p[n]\right)\\
\nonumber ~\text{s.t.}&~ (\ref{Traj_constraint})~\text{and}~(\ref{Altitude_constraint}).
\end{align}

By introducing auxiliary variables $\{{\zeta}_k[n]\}$ and $\{{\eta}_j[n]\}$, problem $(\text{P2-{\uppercase\expandafter{\romannumeral2}}.4})$ is equivalently re-expressed as
\begin{subequations}
\begin{align}
(\text{P2-{\uppercase\expandafter{\romannumeral2}}.5}):&\max\limits_{\{\mv{q}[n],z[n],{\zeta}_k[n],{\eta}_j[n]\}}\frac{1}{N}\sum_{n\in \mathcal{N}}\hat{R}^{(\text{\uppercase\expandafter{\romannumeral2}})}\left({\zeta}_k[n],{\eta}_j[n],p[n]\right)\nonumber\\
\text{s.t.}~&{\zeta}_k[n]\geq \left(||\mv{q}[n]-\mv{w}_{bk}||^2+z^2[n]\right)^{\frac{\alpha}{2}},\forall k\in \mathcal{K}, n\in \mathcal{N}\label{P23_1_C}\\
&{\eta}_j[n]\leq \left(||\mv{q}[n]-\mv{w}_{ej}||^2+z^2[n]\right)^{\frac{\alpha}{2}},\forall j\in \mathcal{J}, n\in \mathcal{N}\label{Nonconvex1_C}\\
&(\ref{Traj_constraint}),~(\ref{Altitude_constraint}),\nonumber
\end{align}
\end{subequations}
where $\hat{R}^{(\text{\uppercase\expandafter{\romannumeral2}})}\left({\zeta}_k[n],{\eta}_j[n],p[n]\right)={\log _2}\left( {1 + \sum\limits_{k \in {\cal K}} {\frac{{{\beta _0}p[n]}}{{{\zeta _k}[n]}}} } \right) - {\log _2}\bigg( {1 + \sum\limits_{j \in {\cal J}} {\frac{{{\beta _0}p[n]}}{{{\eta_j}[n]}}} } \bigg)$.
Notice that the function $\hat{R}^{(\text{\uppercase\expandafter{\romannumeral2}})}\left({\zeta}_k[n],{\eta}_j[n],p[n]\right)$ in the objective function and the RHS terms in constraint (\ref{Nonconvex1_C}) are convex with respect to $\{\mv{q}[n], z[n]\}$, making problem $(\text{P2-{\uppercase\expandafter{\romannumeral2}}.5})$ non-convex.
To deal with this issue, we apply the SCA technique to obtain a converged solution to problem (P2-II.5) in an iterative manner.
At each iteration $m\geq1$, suppose that the local trajectory point is given as $\{\mv q^{(m)}[n], z^{(m)}[n]\}$.
Then, we have the lower bounds for the function $\hat R^{(\text{\uppercase\expandafter{\romannumeral2}})}\left({\zeta}_k[n],{\eta}_j[n],p[n]\right)$ and the RHS term of (\ref{Nonconvex1_C}) as follows based on the first-order Taylor expansion.
\begin{align}
\nonumber&\hat R^{(\text{\uppercase\expandafter{\romannumeral2}})}\left({\zeta}_k[n],{\eta}_j[n],p[n]\right)\ge {{\hat R}^{(\text{\uppercase\expandafter{\romannumeral2}}m)}}\left({\zeta}_k[n],{\eta}_j[n],p[n]\right)
\triangleq {\log _2}\bigg( {1 + \sum\limits_{k \in {\cal K}} {\frac{{{\beta _0}p[n]}}{{\zeta_k^{(m)}[n]}}} } \bigg)\\
&- {\log _2}\bigg( {1 + \sum\limits_{j \in {\cal J}} {\frac{{{\beta _0}p[n]}}{{{\eta_j}[n]}}} } \bigg)
-\frac{1}{\ln2} {{\bigg( {1 + \sum\limits_{k \in {\cal K}} {\frac{{{\beta _0}p[n]}}{{\zeta _k^{(m)}[n]}}} } \bigg)}^{-1}}\sum\limits_{k \in {\cal K}} {\bigg({\frac{{{\beta _0}p[n]}}{{{{\zeta _k^{(m)2}[n]}}}}\left( {{\zeta _k}[n] - \zeta_k^{(m)}[n]} \right)} \bigg)}\label{convex2_C},
\end{align}
\begin{align}\label{convex1_C}
&\left(||\mv{q}[n]-\mv{w}_{ej}||^2+z^{2}[n]\right)^{\alpha/2}\geq\alpha\Big[{z^{(m)}}[n](z[n]- {z^{(m)}}[n]) + ({{\mv{q}}^{(m)}}[n] - {{\mv{w}}_{ej}})({\mv{q}}[n] - {{\mv{q}}^{(m)}}[n])\Big]
\nonumber\\&\times\left(||\mv{q}^{(m)}[n]-\mv{w}_{ej}||^2+z^{(m)2}[n]\right)^{(\alpha-2)/2}\!\!\!
+\left(||\mv{q}^{(m)}[n]-\mv{w}_{ej}||^2+z^{(m)2}[n]\right)^{\alpha/2}\triangleq {E_{ej}^{\text{lb}}}^{(\text{\uppercase\expandafter{\romannumeral2}})}[n].
\end{align}
Replacing $\hat{R}^{(\text{\uppercase\expandafter{\romannumeral2}})}\left({\zeta}_k[n],{\eta}_j[n],p[n]\right)$ and the RHS terms in (\ref{Nonconvex1_C}) as ${{\hat R}^{(\text{\uppercase\expandafter{\romannumeral2}}m)}}\left({\zeta}_k[n],{\eta}_j[n],p[n]\right)$ and ${E_{ej}^{\text{lb}}}^{(\text{\uppercase\expandafter{\romannumeral2}})}[n]$, respectively, problem $(\text{P2-{\uppercase\expandafter{\romannumeral2}}.5})$ is approximately expressed as the following convex optimization problem that can be efficiently solved by CVX \cite{grant2014cvx}.
\begin{align}
\nonumber(\text{P2-\uppercase\expandafter{\romannumeral2}.6}.m):~&\max\limits_{\{ {\mv{q}}[n],z[n],{\zeta_k}[n],{\eta_j}[n]\} } \frac{1}{N}\sum\limits_{n\in \mathcal{N}} {{{\hat R}^{(\text{\uppercase\expandafter{\romannumeral2}}m)}}}\left({\zeta}_k[n],{\eta}_j[n],p[n]\right)\\
\nonumber\text{s.t.}~~&{\eta}_j[n]\leq {E_{ej}^{\text{lb}}}^{(\text{\uppercase\expandafter{\romannumeral2}})}[n],\forall j\in \mathcal{J}, n\in \mathcal{N}\\
\nonumber&(\ref{Traj_constraint}),~(\ref{Altitude_constraint}),~\text{and}~(\ref{P23_1_C}).
\end{align}

Therefore, at iteration $(m+1)$, we update the UAV trajectory point $\{\mv{q}^{(m+1)}[n], z^{(m+1)}[n]\}$ as the optimal solution to the approximate problem (P2{-\uppercase\expandafter{\romannumeral2}}.6.$m$), under the local trajectory point $\{\mv{q}^{(m)}[n], z^{(m)}[n]\}$ in the previous iteration $m$.
As the iteration converges, we can obtain an efficient solution to problem $(\text{P2-{\uppercase\expandafter{\romannumeral2}}.4})$.

To sum up, we solve for the transmit power $\{p[n]\}$ and the trajectory $\{\mv{q}[n], z[n]\}$ in an iterative manner above, and accordingly, we obtain an efficient solution to problem $(\text{P2-{\uppercase\expandafter{\romannumeral2}}.1})$ or equivalently (P2-II).

\section{Numerical Results}\label{nummerical}
In this section, we conduct numerical results to validate the performance of our proposed designs.
In the simulation, unless otherwise stated, we use the following settings:
$K=3$,
$J=2$,
$\mv{w}_{b1} = (-100~\textrm{m},300~\textrm{m})$, $\mv{w}_{b2} = (0~\textrm{m},300~\textrm{m})$, $\mv{w}_{b3} = (100~\textrm{m},300~\textrm{m})$,
$\mv{w}_{e1} = (-50~\textrm{m},180~\textrm{m})$, $\mv{w}_{e2} = (0~\textrm{m},180~\textrm{m})$,
$\mv{q}[0] = (-305~\textrm{m},800~\textrm{m})$, $\mv{q}[N+1] = (-80~\textrm{m},-200~\textrm{m})$,
$z[0] = z[N+1] = 200~\textrm{m}$, $z_\text{min}=150~\textrm{m}$, $z_\text{max}=250~\textrm{m}$,
$t_s=0.5~\textrm{s}$, $\alpha=2$, $\tilde{V}=25~\textrm{m/s}$, $\tilde{V}_{\text{up}}=4~\textrm{m/s}$, $\tilde{V}_{\text{down}}=6~\textrm{m/s}$,
$P=P_\text{ave}=30~\textrm{dBm}$, $P_\text{peak}=4P_\text{ave}$, $\beta_0 = -30~\textrm{dBm}$ and $\sigma^2=-80~\textrm{dBm}$.

\subsection{Quasi-Stationary UAV Scenario}
\begin{figure}[!h]
  \centering
  \subfigure[Case with non-colluding eavesdroppers]{
    \label{fig:subfig:a} 
    \includegraphics[width=6.8cm]{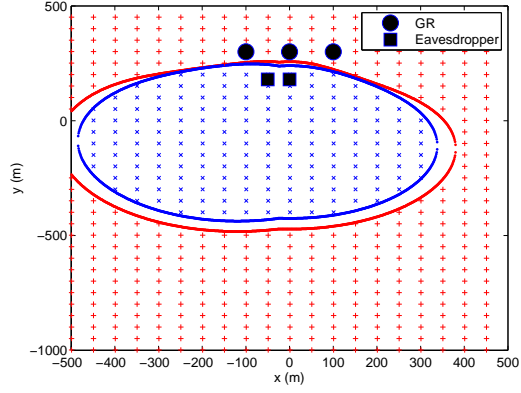}}

  \subfigure[Case with colluding eavesdroppers]{
    \label{fig:subfig:b} 
    \includegraphics[width=6.8cm]{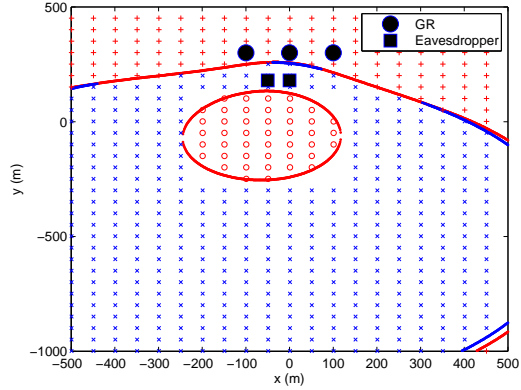}}
  \caption{Obtained optimal UAV altitude to problems (P1-I.2) and (P1-II.2) under different given horizontal location. The region marked with $+$ denotes that the optimal UAV altitude is at the lowest. The region marked with $\times$ denotes that the optimal UAV altitude is at the highest.
  The blank region between the former two regions denotes that the optimal UAV altitude is between the minimum and maximum values.
  The region marked with $\circ$ denotes that the secrecy rate becomes zero, regardless of the UAV altitude.}
  \label{fig:subfig} 
\end{figure}
Figs. \ref{fig:subfig:a} and \ref{fig:subfig:b} show the obtained optimal UAV altitude to problems (P1-I.2) and (P1-II.2) under different given horizontal locations for non-colluding and colluding eavesdroppers.
It is observed that in both figures, when the UAV is located closer to GRs than eavesdroppers, the UAV should stay at the lowest altitude to enjoy the strong legitimate communication link; while when the UAV is located closer to eavesdroppers than GRs with medium distance, the UAV should be deployed at the highest altitude to minimize the information leakage.
It is also observed that between the regions with lowest and highest altitudes, there exists a region where the UAV should stay at an altitude between the lowest and highest values.
Furthermore, it is observed in Fig. \ref{fig:subfig:a} that with the non-colluding eavesdroppers, the secrecy rate is always positive even when the UAV is located close to eavesdroppers. By contrast, it is observed in Fig. \ref{fig:subfig:b} that with the colluding eavesdroppers, there exists a region where the secrecy rate becomes zero due to the collaborative interception by colluding eavesdroppers.

\begin{figure}[!h]
\centering
  \includegraphics[width=6.8cm]{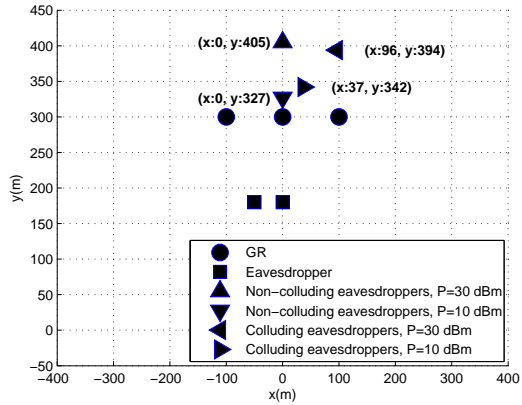}\\
  \caption{Obtained optimal horizontal location of UAV under non-colluding and colluding eavesdroppers.}
  \label{fig:quasi_horizontal.eps}
\end{figure}
Next, we consider problems (P1-I) and (P1-II) when both horizontal and vertical locations of UAV are optimized jointly with the transmit power control.
Fig. \ref{fig:quasi_horizontal.eps} shows the obtained optimal horizontal locations of the UAV under non-colluding and colluding eavesdroppers, where the optimal vertical locations are obtained as $z^{\star} = z^{\star\star} = z_{\text{min}}$ under this setup.
It is observed that under the same value of transmit power levels, the optimal horizontal location under colluding eavesdroppers is farther away from the eavesdroppers than that under non-colluding eavesdroppers, in order to better combat against the stronger collaborative interception.
It is also observed that as the transmit power increases, the optimal horizontal locations move farther away from both GRs and eavesdroppers, to achieve a balance between the desirable signal strength at GRs and undesirable information leakage at eavesdroppers.

\begin{figure}[!h]
\centering
  \includegraphics[width=6.8cm]{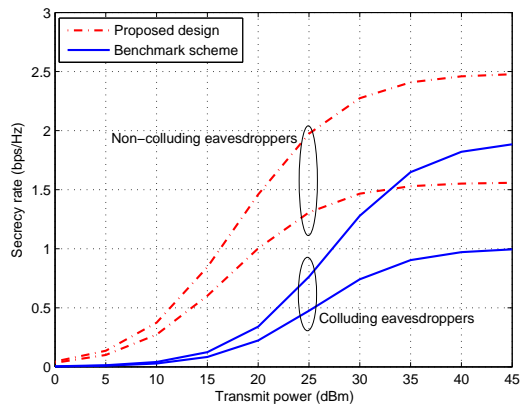}\\
  \caption{Secrecy rate versus the UAV's maximum transmit power under non-colluding and colluding eavesdroppers.}
  \label{fig:quasi_power}
\end{figure}
Fig. \ref{fig:quasi_power} shows the achieved secrecy rate versus the UAV's maximum transmit power under both non-colluding and colluding eavesdroppers.
For comparison, we consider benchmark schemes when the UAV is located at an optimized location with the received SNR at the GRs being maximized.
It is observed that under both non-colluding and colluding eavesdroppers, our proposed designs significantly outperform the benchmark schemes.
It is also observed that the achieved secrecy rate under non-colluding eavesdroppers is always larger than that under colluding eavesdroppers. This is intuitive, as the colluding eavesdropping is more harmful than the non-colluding one.

\subsection{Mobile UAV Scenario}
Next, we show the performance of our proposed designs in the mobile UAV scenario.

%

\begin{figure}[!h]
\begin{minipage}[t]{0.49\linewidth}
\centering
  \includegraphics[width=6.8cm]{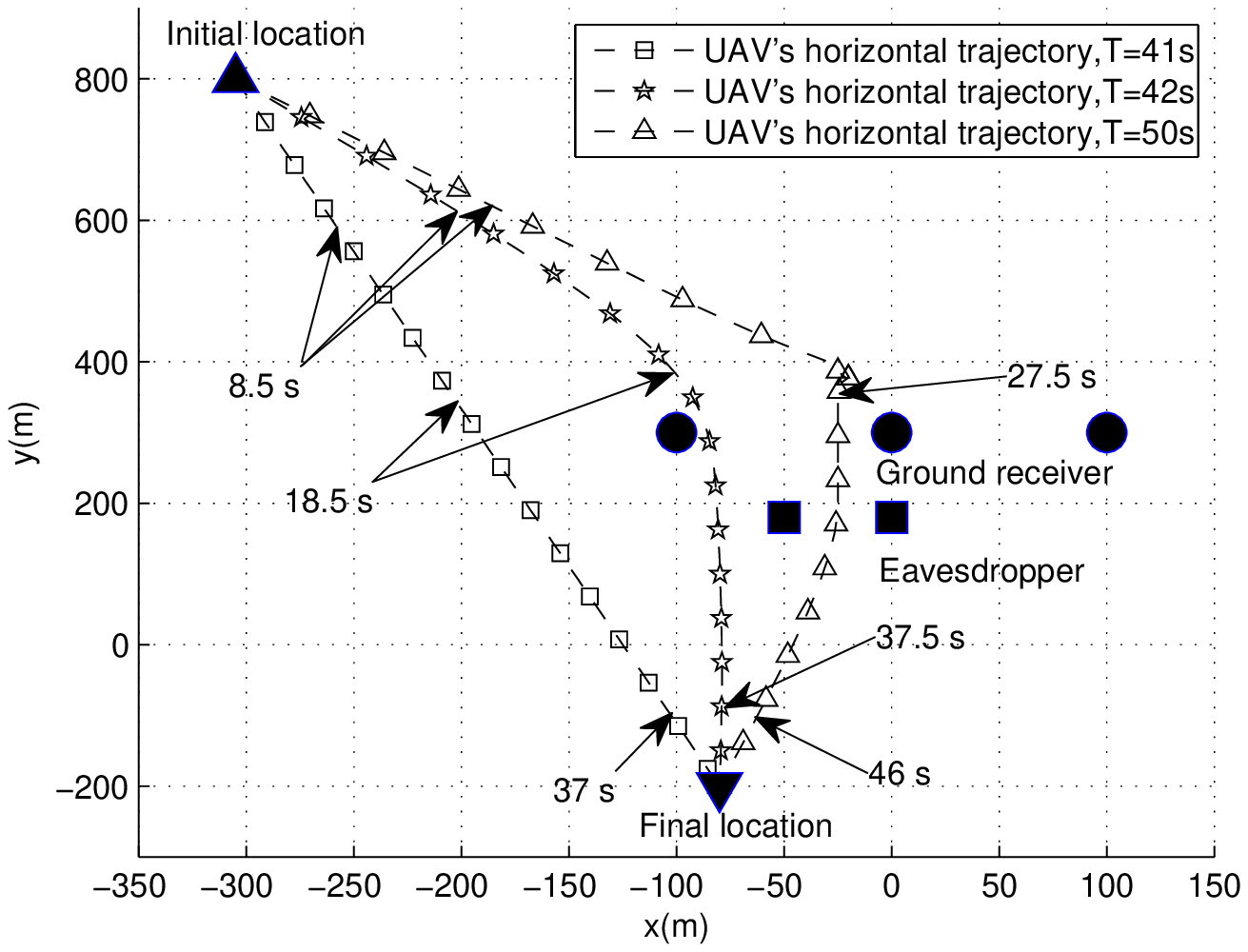}\\
  \caption{Obtained UAV horizontal trajectories by the proposed design under non-colluding eavesdroppers, which are sampled every $2.5$ seconds.}
  \label{fig:trajectory_noncolluding}
\end{minipage}
\hfill
\begin{minipage}[t]{0.49\linewidth}
\centering
  \includegraphics[width=6.8cm]{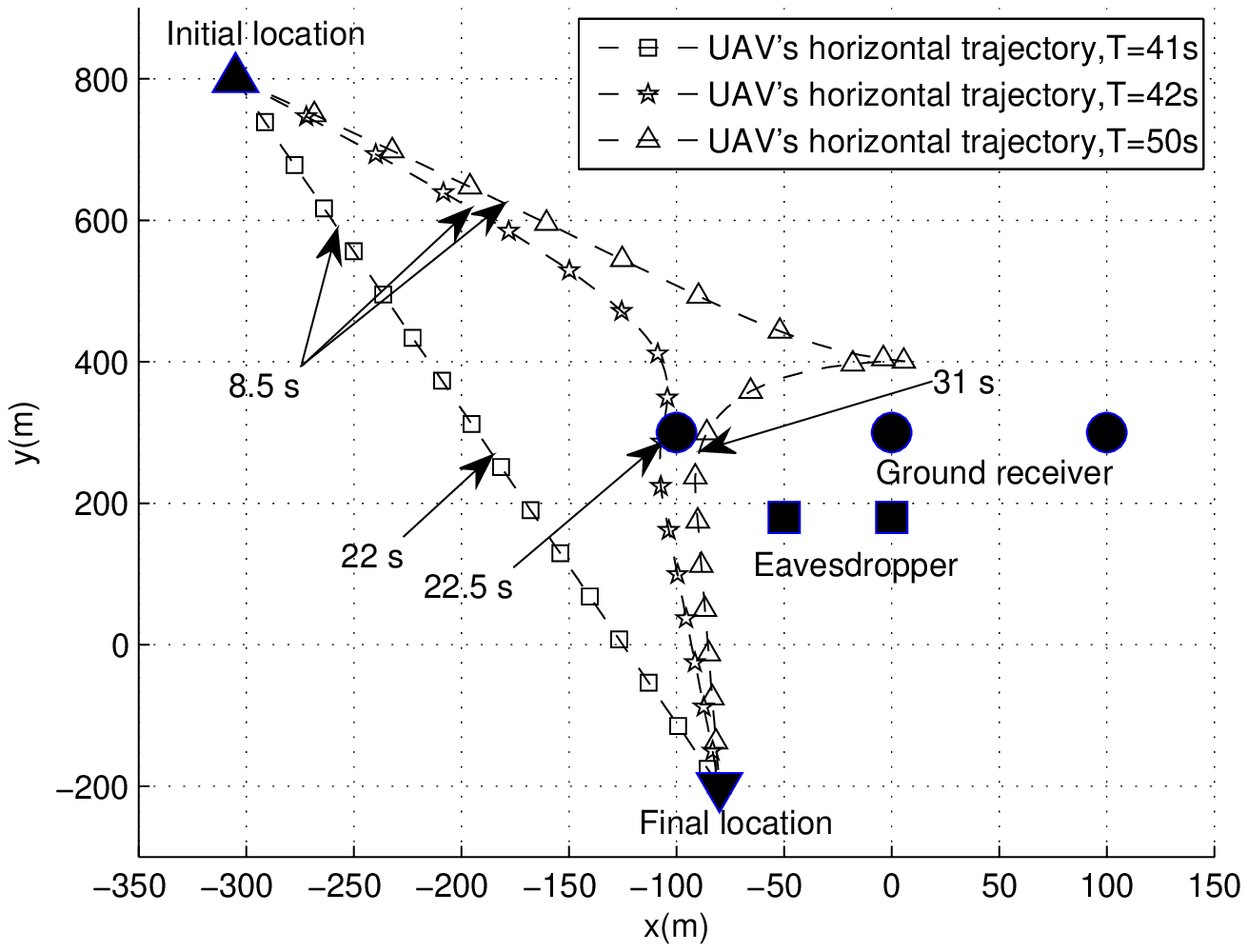}\\
  \caption{Obtained UAV horizontal trajectories by the proposed design under colluding eavesdroppers, which are sampled every $2.5$ seconds.}
  \label{fig:trajectory}
\end{minipage}
\end{figure}
Figs. \ref{fig:trajectory_noncolluding} and \ref{fig:trajectory} show the obtained horizontal trajectories of the UAV for non-colluding and colluding eavesdroppers by our proposed designs, under mission duration $T = 41$ s, $T = 42$ s, and $T = 50$ s, respectively.
When $T$ is large (e.g., $T =50$ s), the UAV is observed to hover at an optimized point with longest duration.
In particular, under non-colluding eavesdroppers, the UAV is observed to fly straightly at the maximum speed from the hovering location to the final location;
while under colluding eavesdroppers, the UAV is observed to fly to the final location following arc paths that are away from the eavesdroppers to avoid the collaborative information interception.
When $T$ is small (e.g., $T = 41$ s and $T = 42$ s), the UAV is observed to fly at the maximum speed towards the hovering location as close as possible, but they cannot exactly reach there due to the time and speed limitations.

%

\begin{figure}[!h]
\begin{minipage}[t]{0.49\linewidth}
\centering
  \includegraphics[width=6.8cm]{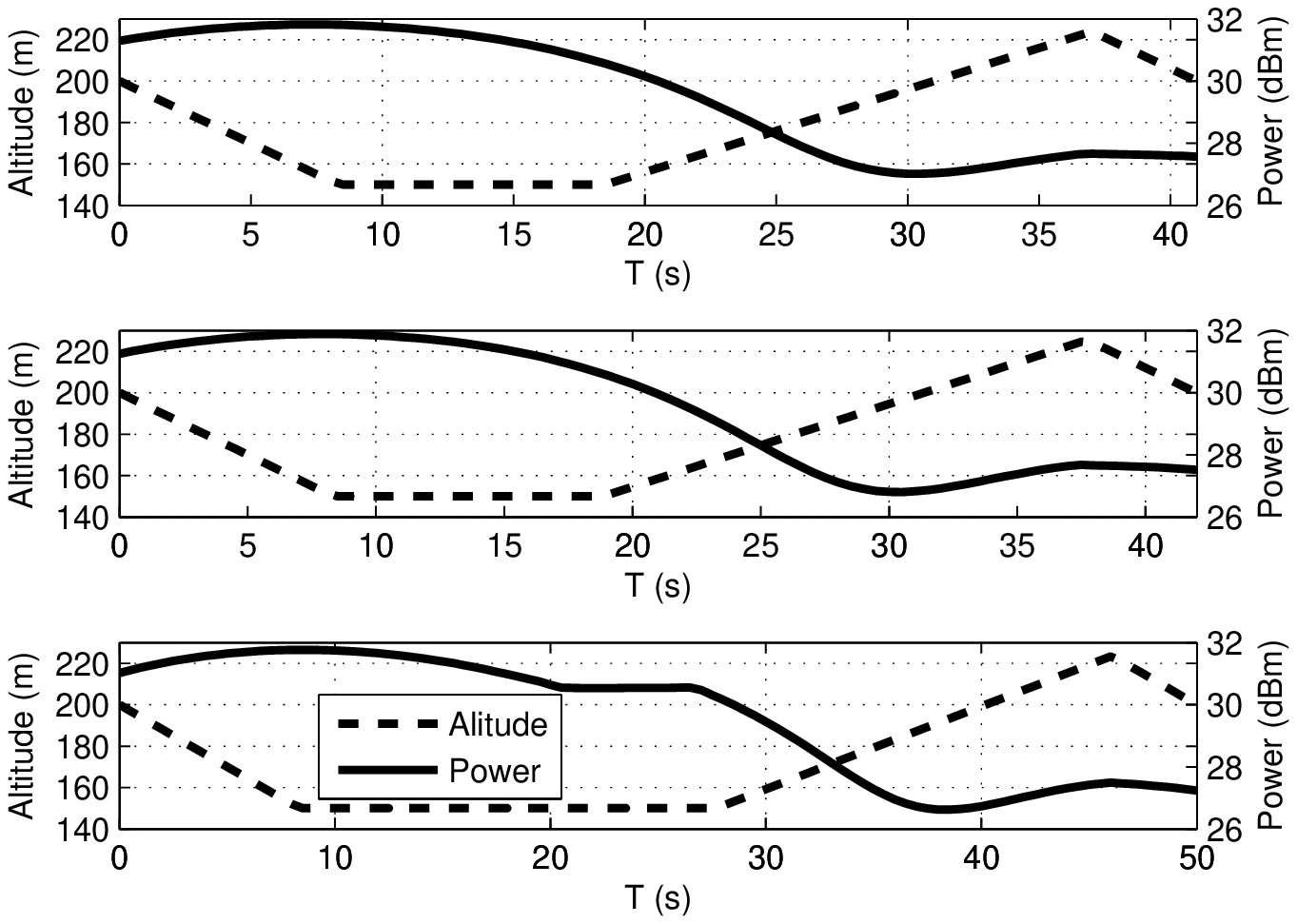}\\
  \caption{Obtained UAV altitude and transmit power over time by our proposed designs under non-colluding eavesdroppers.}
  \label{fig:altitude&power_noncolluding}
\end{minipage}
\hfill
\begin{minipage}[t]{0.49\linewidth}
\centering
  \includegraphics[width=6.8cm]{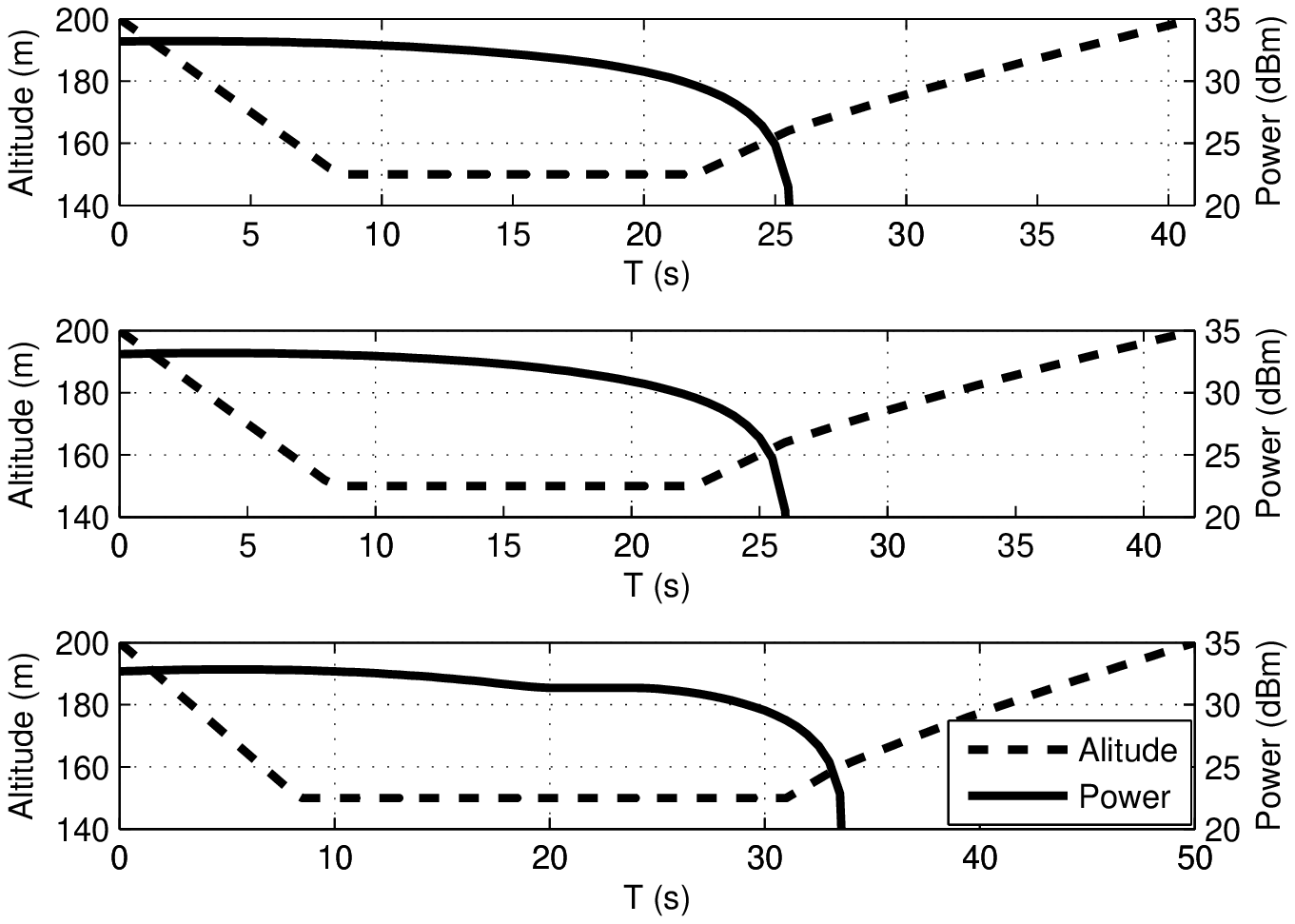}\\
  \caption{Obtained UAV altitude and transmit power over time by our proposed designs under colluding eavesdroppers.}
  \label{fig:altitude&power}
\end{minipage}
\end{figure}

Figs. \ref{fig:altitude&power_noncolluding} and \ref{fig:altitude&power} show the obtained UAV altitude and transmit power under non-colluding and colluding eavesdroppers by our proposed designs, under mission duration $T = 41$ s, $T = 42$ s, and $T = 50$ s, respectively.
Under both non-colluding and colluding eavesdroppers, it is observed that when the UAV is closer to the GRs than the eavesdroppers, it drops its altitude to enhance the desirable information transmission;
by contrast, when UAV is closer to the eavesdroppers than the GRs, it lifts its altitude and decreases the transmit power to prevent the undesirable information leakage.
This observation is generally consistent with our discussion in Remark \ref{Remark_P1.3_solution}, which shows the optimized altitude behavior under given horizontal UAV location.
In particular, when the mission duration is short with $T = 41$ s, it is observed that there is no additional time for the UAV to adjust the its horizontal trajectory, but it still can adjust its altitude for achieving better communication performance.
Furthermore, it is observed that under non-colluding eavesdroppers, the UAV always has positive transmit power (thus positive secrecy rate) during its flight; while under colluding eavesdroppers, the UAV sets its transmit power to be zero (thus leading to zero secrecy rate) at certain points during its flight, especially when the UAV flies close to the eavesdroppers.

%

\begin{figure}[!h]
\begin{minipage}[t]{0.49\linewidth}
\centering
  \includegraphics[width=6.8cm]{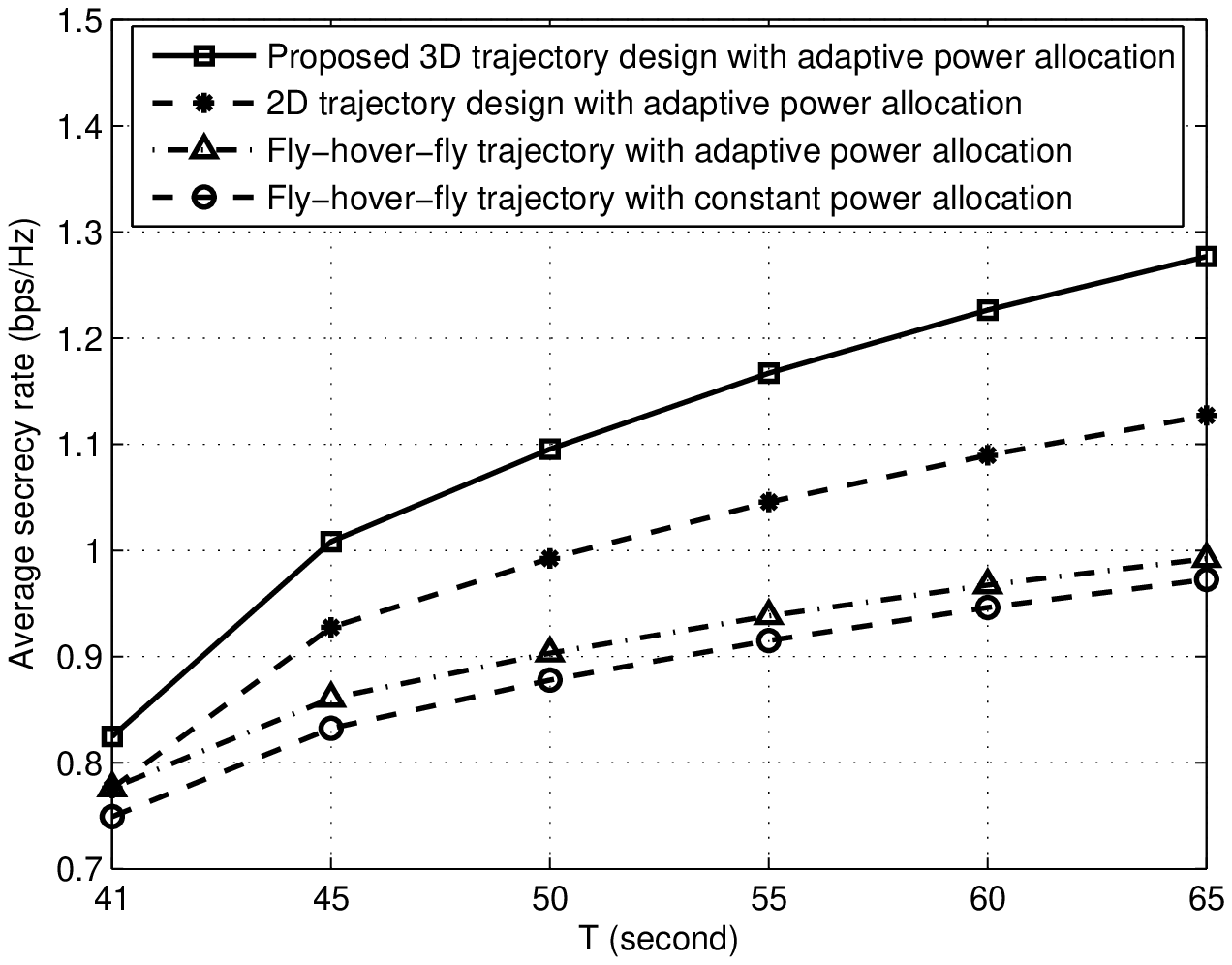}\\
  \caption{Average secrecy rate versus mission duration $T$ under non-colluding eavesdroppers.}
  \label{fig:rate_T_noncolluding}
\end{minipage}
\hfill
\begin{minipage}[t]{0.49\linewidth}
\centering
  \includegraphics[width=6.8cm]{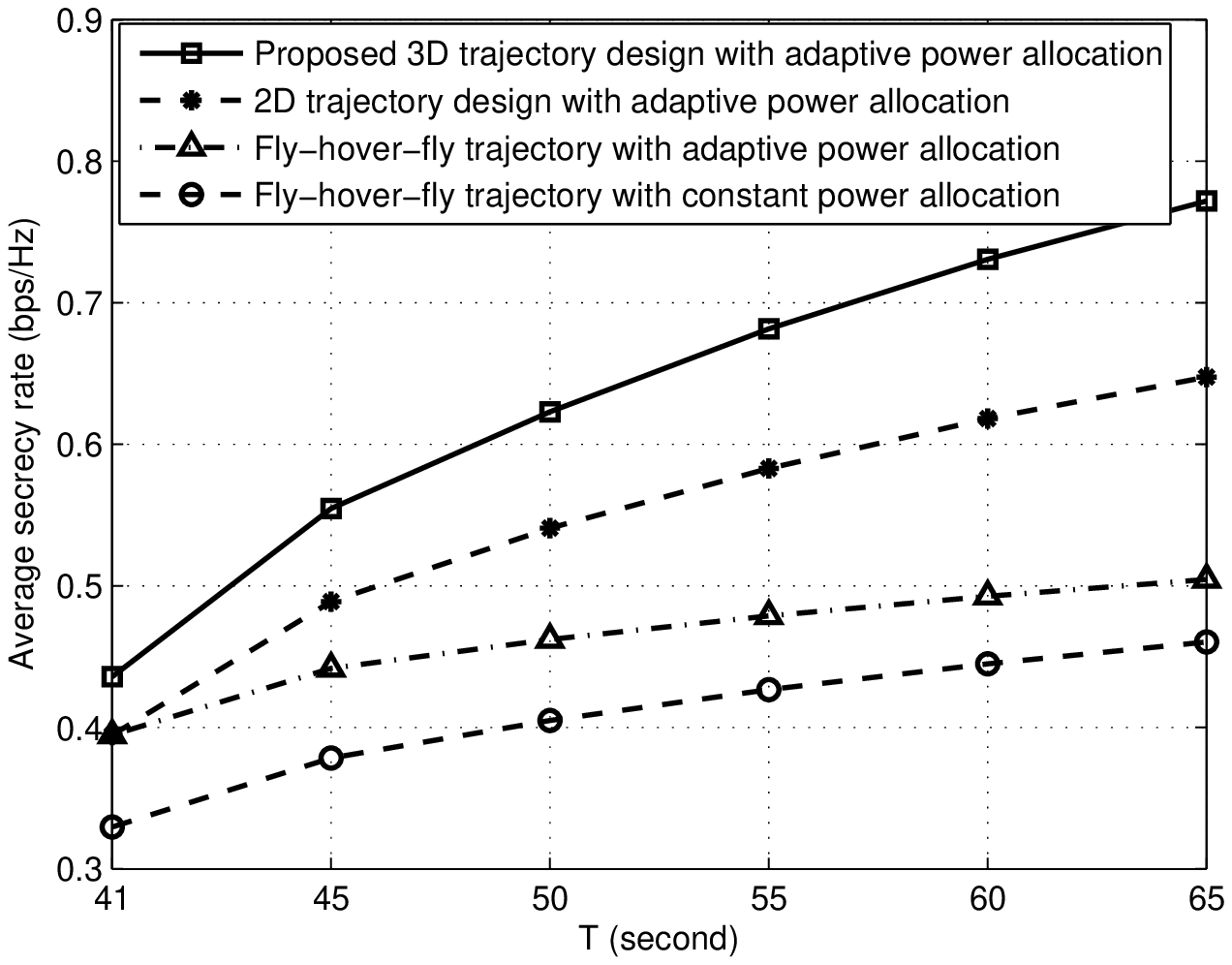}\\
  \caption{Average secrecy rate versus mission duration $T$ under colluding eavesdroppers.}
  \label{fig:rate}
\end{minipage}
\end{figure}

Figs. \ref{fig:rate_T_noncolluding} and \ref{fig:rate} show the average secrecy rate versus the mission duration $T$ for non-colluding and colluding eavesdroppers, respectively.
For comparison, we consider the following three benchmark schemes.
\begin{itemize}
\item {\bf 2D trajectory design with adaptive power allocation}: The UAV flies at a fixed altitude $z$, and jointly optimizes its 2D horizontal trajectory and the transmit power allocation to maximize the average secrecy rate. This design can be implemented by solving problem $(\text{P2-{\uppercase\expandafter{\romannumeral1}}})$ or $(\text{P2-{\uppercase\expandafter{\romannumeral2}}})$ under given $z$. For this design, we set the fixed altitude as $z = 200$ m.

\item {\bf Fly-hover-fly trajectory with adaptive power allocation}: The UAV adopts the fly-hover-fly trajectory in Remark \ref{remark1}, during which it adaptively optimizes the transmit power allocation by solving problem $(\text{P2-{\uppercase\expandafter{\romannumeral1}}.2})$ or $(\text{P2-{\uppercase\expandafter{\romannumeral2}}.2})$.

\item {\bf Fly-hover-fly trajectory with constant power allocation}: The UAV adopts the fly-hover-fly trajectory in Remark \ref{remark1}, during which it employs the constant power allocation, i.e., $p[n] = P_\textrm{ave}, \forall n\in\mathcal{N}$.
\end{itemize}
It is observed that as $T$ increases, the secrecy rates achieved by all the four schemes increase under both non-colluding and colluding eavesdroppers. This is due to the fact that in this case, the UAV can fly closer towards the hovering location and/or hover there with longer duration, thus leading to higher average achievable secrecy rate.
When $T$ is small (e.g. $T = 40$ s), the 2D trajectory design with adaptive power allocation is observed to have a similar performance as the fly-hover-fly trajectory with adaptive power allocation.
This is due to the fact that in this case, the mission duration is only sufficient for the UAV to fly from the initial to final locations, and there is no additional time to adjust the trajectory for communication performance optimization.
When $T$ becomes large, the two schemes of 3D and 2D trajectory design with adaptive power allocation are observed to significantly outperform the other two benchmark schemes with fly-hover-fly trajectory under both non-colluding and colluding eavesdroppers scenarios.
Furthermore, it is observed that there is a large performance gap between the proposed 3D trajectory design with adaptive power allocation versus the 2D trajectory design with adaptive power allocation.
This shows the significance of adapting the vertical UAV trajectory or altitude in enhancing the secrecy UAV communication performance with CoMP reception.

\section{Conclusion}
In this paper, we considered the CoMP reception-enabled secrecy UAV communication system, in which multiple GRs cooperatively detect the legitimate information sent from the UAV to enhance the legitimate communication performance under quasi-stationary and mobile UAVs scenarios.
By considering both non-colluding and colluding eavesdroppers under both scenarios, we proposed to jointly exploit the UAV's 3D maneuver and transmit power adaptation to maximize the average secrecy rate.
It is shown that the joint 3D maneuver and transmit power optimization greatly enhances the secrecy performance, as compared to other benchmark schemes with e.g. 2D maneuver optimization only.
It is also shown that the 3D maneuver behaviors under non-colluding and colluding eavesdropper scenarios are distinct.
How to extend the results to other scenarios, e.g., with multiple UAVs and multi-antenna GRs, are interesting directions worth further investigation.

\footnotesize
\bibliographystyle{IEEEtran}
\bibliography{myreference}

\begin{thebibliography}{10}
\providecommand{\url}[1]{#1}
\csname url@samestyle\endcsname
\providecommand{\newblock}{\relax}
\providecommand{\bibinfo}[2]{#2}
\providecommand{\BIBentrySTDinterwordspacing}{\spaceskip=0pt\relax}
\providecommand{\BIBentryALTinterwordstretchfactor}{4}
\providecommand{\BIBentryALTinterwordspacing}{\spaceskip=\fontdimen2\font plus
\BIBentryALTinterwordstretchfactor\fontdimen3\font minus
  \fontdimen4\font\relax}
\providecommand{\BIBforeignlanguage}[2]{{%
\expandafter\ifx\csname l@#1\endcsname\relax
\typeout{** WARNING: IEEEtran.bst: No hyphenation pattern has been}%
\typeout{** loaded for the language `#1'. Using the pattern for}%
\typeout{** the default language instead.}%
\else
\language=\csname l@#1\endcsname
\fi
#2}}
\providecommand{\BIBdecl}{\relax}
\BIBdecl

\bibitem{yao3D2019}
J.~Yao, C.~Zhong, Z.~Liu, and J.~Xu, ``{3D} trajectory optimization for secure
  {UAV} communication with {CoMP} reception,'' in \emph{Proc. IEEE Global
  Commun. Conf. (GLOBECOM)}, Dec. 2019, pp. 1--6.

\bibitem{ZengAccessing2019}
Y.~{Zeng}, Q.~{Wu}, and R.~{Zhang}, ``Accessing from the sky: A tutorial on
  {UAV} communications for {5G} and beyond,'' \emph{Proc. IEEE}, vol. 107,
  no.~12, pp. 2327--2375, Dec. 2019.

\bibitem{ZengCellular2019}
Y.~{Zeng}, J.~{Lyu}, and R.~{Zhang}, ``Cellular-connected {UAV}: Potential,
  challenges, and promising technologies,'' \emph{IEEE Wireless Commun.},
  vol.~26, no.~1, pp. 120--127, Feb. 2019.

\bibitem{ZhangCellular2019}
S.~Zhang, Y.~Zeng, and R.~Zhang, ``Cellular-enabled {UAV} communication: A
  connectivity-constrained trajectory optimization perspective,'' \emph{IEEE
  Trans. Commun.}, vol.~67, no.~3, pp. 2580--2604, Mar. 2019.

\bibitem{MozaffariBeyond2019}
M.~{Mozaffari}, A.~T.~Z. {Kasgari}, W.~{Saad}, M.~{Bennis}, and M.~{Debbah},
  ``Beyond {5G} with {UAVs}: Foundations of a {3D} wireless cellular network,''
  \emph{IEEE Trans. Wireless Commun.}, vol.~18, no.~1, pp. 357--372, Jan. 2019.

\bibitem{MenouarUAV2017}
H.~Menouar, I.~Guvenc, K.~Akkaya, A.~S. Uluagac, A.~Kadri, and A.~Tuncer,
  ``{UAV}-enabled intelligent transportation systems for the smart city:
  Applications and challenges,'' \emph{IEEE Commun. Mag.}, vol.~55, no.~3, pp.
  22--28, Mar. 2017.

\bibitem{XiaoEnabling2016}
Z.~Xiao, P.~Xia, and X.~Xia, ``Enabling {UAV} cellular with millimeter-wave
  communication: Potentials and approaches,'' \emph{IEEE Commun. Mag.},
  vol.~54, no.~5, pp. 66--73, May 2016.

\bibitem{ZengWireless2016}
Y.~Zeng, R.~Zhang, and T.~J. Lim, ``Wireless communications with unmanned
  aerial vehicles: Opportunities and challenges,'' \emph{IEEE Commun. Mag.},
  vol.~54, no.~5, pp. 36--42, May 2016.

\bibitem{XuUAV2018}
J.~Xu, Y.~Zeng, and R.~Zhang, ``{UAV}-enabled wireless power transfer:
  Trajectory design and energy optimization,'' \emph{IEEE Trans. Wireless
  Commun.}, vol.~17, no.~8, pp. 5092--5106, Aug. 2018.

\bibitem{LiPlacement2018}
P.~Li and J.~Xu, ``Placement optimization for {UAV}-enabled wireless networks
  with multi-hop backhauls,'' \emph{J. Commun. Inf. Networks}, vol.~3, no.~4,
  pp. 64--73, Dec. 2018.

\bibitem{XieThroughput2019}
L.~{Xie}, J.~{Xu}, and R.~{Zhang}, ``Throughput maximization for {UAV}-enabled
  wireless powered communication networks,'' \emph{IEEE Internet Things J.},
  vol.~6, no.~2, pp. 1690--1703, Apr. 2019.

\bibitem{EsrafilianLearning2019}
O.~{Esrafilian}, R.~{Gangula}, and D.~{Gesbert}, ``Learning to communicate in
  {UAV}-aided wireless networks: Map-based approaches,'' \emph{IEEE Internet
  Things J.}, vol.~6, no.~2, pp. 1791--1802, Apr. 2019.

\bibitem{ChenEfficient2019}
J.~{Chen} and D.~{Gesbert}, ``Efficient local map search algorithms for the
  placement of flying relays,'' \emph{IEEE Trans. Wireless Commun.}, pp. 1--1,
  2019.

\bibitem{BerghLTE2016}
B.~V.~D. Bergh, A.~Chiumento, and S.~Pollin, ``{LTE} in the sky: Trading off
  propagation benefits with interference costs for aerial nodes,'' \emph{IEEE
  Commun. Mag.}, vol.~54, no.~5, pp. 44--50, May 2016.

\bibitem{YalinizThe2016}
I.~Bor-Yaliniz and H.~Yanikomeroglu, ``The new frontier in {RAN} heterogeneity:
  Multi-tier drone-cells,'' \emph{IEEE Commun. Mag.}, vol.~54, no.~11, pp.
  48--55, Nov. 2016.

\bibitem{YaoSecure2016}
J.~Yao, S.~Feng, X.~Zhou, and Y.~Liu, ``Secure routing in multihop wireless
  ad-hoc networks with decode-and-forward relaying,'' \emph{IEEE Trans.
  Commun.}, vol.~64, no.~2, pp. 753--764, Feb. 2016.

\bibitem{YanSecret2018}
S.~Yan, X.~Zhou, N.~Yang, T.~D. Abhayapala, and A.~L. Swindlehurst, ``Secret
  channel training to enhance physical layer security with a full-duplex
  receiver,'' \emph{IEEE Trans. Inf. Foren. Sec.}, vol.~13, no.~11, pp.
  2788--2800, Nov. 2018.

\bibitem{Yao2019secrecy}
J.~{Yao} and J.~{Xu}, ``Secrecy transmission in large-scale {UAV}-enabled
  wireless networks,'' \emph{IEEE Trans. Commun.}, vol.~67, no.~11, pp.
  7656--7671, Nov. 2019.

\bibitem{ZhuSecrecy2018}
Y.~Zhu, G.~Zheng, and M.~Fitch, ``Secrecy rate analysis of {UAV}-enabled
  mm{W}ave networks using {M}at\'{e}rn hardcore point processes,'' \emph{IEEE
  J. Sel. Areas Commun.}, vol.~36, no.~7, pp. 1397--1409, Jul. 2018.

\bibitem{TangSecrecy2019}
J.~{Tang}, G.~{Chen}, and J.~P. {Coon}, ``Secrecy performance analysis of
  wireless communications in the presence of {UAV} jammer and randomly located
  {UAV} eavesdroppers,'' \emph{IEEE Trans. Inf. Forensics Security}, vol.~14,
  no.~11, pp. 3026--3041, Nov. 2019.

\bibitem{CuiRobust2018}
M.~Cui, G.~Zhang, Q.~Wu, and D.~W.~K. Ng, ``Robust trajectory and transmit
  power design for secure {UAV} communications,'' \emph{IEEE Trans. Veh.
  Technol.}, vol.~67, no.~9, pp. 9042--9046, Sep. 2018.

\bibitem{ZhangSecuring2018}
G.~Zhang, Q.~Wu, M.~Cui, and R.~Zhang, ``Securing {UAV} communications via
  joint trajectory and power control,'' \emph{IEEE Trans. Wireless Commun.},
  vol.~18, no.~2, pp. 1376--1389, Feb. 2019.

\bibitem{WangImproving2017}
Q.~Wang, Z.~Chen, W.~Mei, and J.~Fang, ``Improving physical layer security
  using {UAV}-enabled mobile relaying,'' \emph{IEEE Wireless Commun. Lett.},
  vol.~6, no.~3, pp. 310--313, Jun. 2017.

\bibitem{ZhouSecrecy2017}
Y.~Zhou, P.~L. Yeoh, H.~Chen, Y.~Li, W.~Hardjawana, and B.~Vucetic, ``Secrecy
  outage probability and jamming coverage of {UAV}-enabled friendly jammer,''
  in \emph{Proc. 11th IEEE Australia Int. Conf. Signal Process. Commun. Syst.
  (ICSPCS)}, Dec. 2017, pp. 1--6.

\bibitem{LiRobust2018}
A.~Li, Q.~Wu, and R.~Zhang, ``{UAV}-enabled cooperative jamming for improving
  secrecy of ground wiretap channel,'' \emph{IEEE Wireless Commun. Lett.},
  vol.~8, no.~1, pp. 181--184, Feb. 2019.

\bibitem{BaiEnergy2019}
T.~{Bai}, J.~{Wang}, Y.~{Ren}, and L.~{Hanzo}, ``Energy-efficient computation
  offloading for secure {UAV}-edge-computing systems,'' \emph{IEEE Trans. Veh.
  Technol.}, vol.~68, no.~6, pp. 6074--6087, Jun. 2019.

\bibitem{LeeUAV2018}
H.~Lee, S.~Eom, J.~Park, and I.~Lee, ``{UAV}-aided secure communications with
  cooperative jamming,'' \emph{IEEE Trans. Veh. Technol.}, vol.~67, no.~10, pp.
  9385--9392, Oct. 2018.

\bibitem{ZhongSecure2018}
C.~Zhong, J.~Yao, and J.~Xu, ``Secure {UAV} communication with cooperative
  jamming and trajectory control,'' \emph{IEEE Commun. Lett.}, vol.~23, no.~2,
  pp. 286--289, Feb. 2019.

\bibitem{CaiDual2018}
Y.~Cai, F.~Cui, Q.~Shi, M.~Zhao, and G.~Y. Li, ``Dual-{UAV}-enabled secure
  communications: Joint trajectory design and user scheduling,'' \emph{IEEE J.
  Sel. Areas Commun.}, vol.~36, no.~9, pp. 1972--1985, Sep. 2018.

\bibitem{ZhaoCaching2018}
N.~Zhao, F.~Cheng, F.~R. Yu, J.~Tang, Y.~Chen, G.~Gui, and H.~Sari, ``Caching
  {UAV} assisted secure transmission in hyper-dense networks based on
  interference alignment,'' \emph{IEEE Trans. Commun.}, vol.~66, no.~5, pp.
  2281--2294, May 2018.

\bibitem{ZhouUAV2019}
X.~{Zhou}, Q.~{Wu}, S.~{Yan}, F.~{Shu}, and J.~{Li}, ``{UAV}-enabled secure
  communications: Joint trajectory and transmit power optimization,''
  \emph{IEEE Trans. Veh. Technol.}, vol.~68, no.~4, pp. 4069--4073, Apr. 2019.

\bibitem{SawahashiCoordinated2010}
M.~{Sawahashi}, Y.~{Kishiyama}, A.~{Morimoto}, D.~{Nishikawa}, and M.~{Tanno},
  ``Coordinated multipoint transmission/reception techniques for
  {LTE}-advanced,'' \emph{IEEE Wireless Commun.}, vol.~17, no.~3, pp. 26--34,
  Jun. 2010.

\bibitem{CheckoCloud2015}
A.~{Checko}, H.~L. {Christiansen}, Y.~{Yan}, L.~{Scolari}, G.~{Kardaras}, M.~S.
  {Berger}, and L.~{Dittmann}, ``Cloud {RAN} for mobile networks--{A}
  technology overview,'' \emph{IEEE Commun. Surveys Tuts.}, vol.~17, no.~1, pp.
  405--426, First Quarter 2015.

\bibitem{SimeoneCloud2016}
O.~{Simeone}, A.~{Maeder}, M.~{Peng}, O.~{Sahin}, and W.~{Yu}, ``Cloud radio
  access network: Virtualizing wireless access for dense heterogeneous
  systems,'' \emph{J. Commun. Networks}, vol.~18, no.~2, pp. 135--149, Apr.
  2016.

\bibitem{LiuCoMP2019}
L.~{Liu}, S.~{Zhang}, and R.~{Zhang}, ``{CoMP} in the sky: {UAV} placement and
  movement optimization for multi-user communications,'' \emph{IEEE Trans.
  Commun.}, vol.~67, no.~8, pp. 5645--5658, Aug. 2019.

\bibitem{NgSecure2015}
D.~W.~K. {Ng} and R.~{Schober}, ``Secure and green {SWIPT} in distributed
  antenna networks with limited backhaul capacity,'' \emph{IEEE Trans. Wireless
  Commun.}, vol.~14, no.~9, pp. 5082--5097, Sep. 2015.

\bibitem{KalantariOn2016}
E.~{Kalantari}, H.~{Yanikomeroglu}, and A.~{Yongacoglu}, ``On the number and
  {3D} placement of drone base stations in wireless cellular networks,'' in
  \emph{Proc. IEEE 84th VTC-Fall}, Sep. 2016, pp. 1--6.

\bibitem{Alzenad3D2017}
M.~{Alzenad}, A.~{El-Keyi}, F.~{Lagum}, and H.~{Yanikomeroglu}, ``{3-D}
  placement of an unmanned aerial vehicle base station ({UAV-BS}) for
  energy-efficient maximal coverage,'' \emph{IEEE Wireless Commun. Lett.},
  vol.~6, no.~4, pp. 434--437, Aug. 2017.

\bibitem{SunOptimal2019}
Y.~{Sun}, D.~{Xu}, D.~W.~K. {Ng}, L.~{Dai}, and R.~{Schober}, ``Optimal
  {3D}-trajectory design and resource allocation for solar-powered {UAV}
  communication systems,'' \emph{IEEE Trans. Commun.}, vol.~67, no.~6, pp.
  4281--4298, Jun. 2019.

\bibitem{HuangCognitive2019}
Y.~{Huang}, W.~{Mei}, J.~{Xu}, L.~{Qiu}, and R.~{Zhang}, ``Cognitive {UAV}
  communication via joint maneuver and power control,'' \emph{IEEE Trans.
  Commun.}, vol.~67, no.~11, pp. 7872--7888, Nov. 2019.

\bibitem{KapetanovicPhysical2015}
D.~{Kapetanovic}, G.~{Zheng}, and F.~{Rusek}, ``Physical layer security for
  massive {MIMO}: An overview on passive eavesdropping and active attacks,''
  \emph{IEEE Commun. Mag.}, vol.~53, no.~6, pp. 21--27, Jun. 2015.

\bibitem{Mukherjee2012}
A.~Mukherjee and A.~L. Swindlehurst, ``Detecting passive eavesdroppers in the
  {MIMO} wiretap channel,'' in \emph{Proc. IEEE Int. Conf. Acoustics, Speech,
  and Signal Process.}, Mar. 2012, pp. 2809--2812.

\bibitem{LiangSecure2008}
Y.~{Liang}, H.~V. {Poor}, and S.~{Shamai}, ``Secure communication over fading
  channels,'' \emph{IEEE Trans. Inf. Theory}, vol.~54, no.~6, pp. 2470--2492,
  Jun. 2008.

\bibitem{boyd2004convex}
S.~Boyd and L.~Vandenberghe, \emph{Convex optimization}.\hskip 1em plus 0.5em
  minus 0.4em\relax Cambridge university press, 2004.

\bibitem{grant2014cvx}
\BIBentryALTinterwordspacing
M.~Grant and S.~Boyd, ``{CVX}: Matlab software for disciplined convex
  programming,'' 2016. [Online]. Available: \url{https://cvxr.com/cvx}
\BIBentrySTDinterwordspacing

\bibitem{bertsekas1999nonlinear}
D.~P. Bertsekas, \emph{Nonlinear Programming}.\hskip 1em plus 0.5em minus
  0.4em\relax Belmont, MA, USA: Athena scientific, 1999.

\end{thebibliography}
\end{document}